\documentclass[onecolumn]{aastex6}

\usepackage{color}
\usepackage{hyperref} 

\slugcomment{The Astrophysical Journal, in press}
\shorttitle{Stream-Disk Interaction} 
\shortauthors{Godon}
\watermark{Accepted version}
\setwatermarkfontsize{1.0in}

\begin{document}
\bibliographystyle{apj}

\title{{\bf 
L1 Stream Deflection and Ballistic Launching at the Disk Bow Shock:  
An Absorption-line Velocity Analysis 
in Semi-Detached Binaries   
}} 

\author{Patrick Godon\footnote{Visiting in the Henry A. Rowland 
Department of Physics \& Astronomy at the Johns Hopkins University,
Baltimore, MD 21218, USA}}
\affil{Department of Astrophysics \& Planetary Science,  
Villanova University,
Villanova, PA 19085, USA}
\email{patrick.godon@villanova.edu }

\begin{abstract}

Observations of semi-detached interacting binaries reveal orbital modulation 
in the optical, UV, and X-ray bands, indicating the presence of 
absorbing material obscuring the disk and accreting primary star  
at specific orbital phases consistent with $L1$ stream material
overflowing the disk edge.   

We simulate the L1 stream interaction with the disk using tests particles
within the context of the Roche model in the restricted three-body problem.
At the disk bow shock the L1 stream particles are deflected and launched 
onto ballistic trajectories above the disk (as would normally occurs at  
the front of a detached shock in a hypersonic flow past a blunt body). 
At a given scale height, the material is assumed 
to continue without being affected by the disk, while at lower altitude 
it is being launched at an increasing elevation, as well as
gradually being dragged by the Keplerian flow.  
Near the disk mid-plane ($z << H$) the material is assumed to become part 
of the disk.  
We follow the stream material ballistic trajectories over the disk surface,  
where they reach a maximum height $z/r$   
at a binary phase $\Phi \sim 0.75$, and land onto the disk at a smaller 
radius around phase $\Phi \sim 0.5$.  The phase of the maximum height, 
phase of the landing site and phase of the hot spot itself, all decrease 
significantly with decreasing disk radius.   

The radial velocity for each $L1$ stream ballistic trajectory 
along the line of sight (of the observer) to the hot inner parts of the disk is computed 
as a function of the orbital phase  for a binary configuration matching 
the dwarf nova U Geminorum. 
The computed velocity amplitudes, phases, and pattern
match the observed velocity offsets of the metal lines in the 
{\it FUSE} spectrum of U Gem during outburst. 
As ballistic trajectories are much easier to compute than realistic
three-dimensional hydrodynamical simulations,  
we propose the use of the $L1$ stream deflection and ballistic launching 
as a means for the analysis of the absorption-line orbital variability  
in semi-detached binaries and to assess or confirm, 
with some limitations,
system parameters such as the mass ratio, inclination, and disk outer radius.  

\end{abstract}


\section*{ }  

\clearpage 

\section{{\bf Introduction}} 

Semi-detached interacting binaries are systems          
in which one star, the donor (or secondary), 
fills its Roche lobe and loses matter to the other star, 
the accretor (or primary). 
In most cases, the matter forms a disk around the primary, and as 
it is being accreted (either sporadically, periodically, or continuously), 
highly energetic emission is released. The accreting star can be either 
a main-sequence star (e.g. such as in Algol variables),
a white dwarf (as in cataclysmic variables - CVs),
or even a neutron star or a black hole (as in low-mass x-ray binaries - LMXBs).
Semi-detached binaries have been extensively studied  and
analyzed for many decades now, and they have become the 
favorite objects to observe for the study of accretion disks 
\citep{nus94}.  

In these systems, the secondary reaches its Roche limit and, consequently, 
funnels matter through the first Lagrangian point ($L1$)
into the Roche lobe of the primary \citep{kui41,kru66,pre74}.  
After the stream of matter leaves the $L1$ vicinity, 
it sets onto a ballistic trajectory around the primary,  
and then collides with itself to form a
ring of matter \citep{lub75}, which eventually grows into an accretion disk 
\citep[e.g.][]{mur96}.   
The incoming $L1$ stream continues to interact with the accretion 
disk after its formation by impacting its edge to form a ``hot spot'' 
on the rim  of the disk, 
which appears as an orbital modulation in the light curve that  
cannot be attributed to primary or secondary eclipses.
For example, \citet{krz65} showed that  
the optical light curve of the dwarf nova U Geminorum (a CV) 
exhibits eclipses in quiescence (low mass accretion rate),   
which occur earlier during rise to outburst, and return 
to their  original phase as the systems fades back into quiescence. 
These can be explained as eclipses of the hot spot by the secondary star, 
shifting phase as the hot spot moves along the rim of the disk 
which expands during rise to outburst and shrinks during decline. 

However, a second hot spot often appears on the surface 
of the disk \citep[in Doppler tomograms;][]{mar85,mar88}, 
which is due (as first demonstrated by \citet{lub89}) to the 
$L1$ stream material flowing over the disk edge  
and impacting the disk at smaller radii. 
Theoretically, one can differentiate between two cases of stream overflow.  
In the first case \citep{lub75,lub76,lub89}, 
parts of the $L1$ stream is high enough
above the disk (i.e. where the stream is much denser than the disk) to 
continue on a ballistic trajectory unaffected by the disk edge, and 
impacts the disk near the {\it closest approach radius} 
($\varpi_{\rm min}$ in \citet{lub75}) at phase $\Phi \approx 0.6$.  
In the second case, 
as shown with the advance of three-dimensional hydrodynamical simulations 
\citep[e.g.][]{arm96,blo98,kun01,bis00,bis03},  
the $L1$ stream material is deflected vertically from the impinging 
region (i.e. at the first hot spot) and flies in a more or less diffuse stream 
to inner parts of the disk, hitting the disk surface 
close to the {\it circularization radius} 
($\varpi_{\rm d}$ in \citet{lub75}) at orbital phase $\Phi=0.5$.  

The vertical deflection of stream matter at the (first) hot spot
can itself be due to either deflection by the bow shock itself
\citep[][similar to the isothermal case in \citet{arm98}]{kun01}, 
or due to high pressure of the hot post-shock expanding gas out of
hydrostatic balance in the hot spot region
\citep[][similar to the adiabatic case of \citet{arm98}, and as
observed in WZ Sge by \citet{spr98}]{kun01}.  

These 3D simulations of the stream overflowing the disk have been able 
to explain the veiling of the accreting star 
in X-ray observations of CVs and LMXBs around orbital 
phase $\Phi=0.7$, when the inclination is at least $65^{\circ}$ \citep*{kun01}, 
and can even explain jets perpendicular to the 
orbital plane as observed in Algol-type binaries.   

The complexity of the stream-disk interaction and how this affects the 
observations depend on the system parameters (e.g. such as inclination, 
binary mass ratio) and the state of the system (high mass transfer  
rate versus low mass transfer rate). 
Whether parts of the stream overflow the disk in a ballistic trajectory 
also depends on the parameters such as the disk radius or 
whether the disk and stream are aligned (tilted disks would certainly  
facilitates ballistic stream overflow as the stream may miss the 
disk).  
Also, if the material gains energy, e.g. is accelerated by the disk Keplerian
flow, then it could be ejected or launched into a circumbinary orbit.
It follows that for each system a slightly different model might be needed to explain 
the observations. 

In LMXBs a model of the 
stream-disk interaction was suggested \citep*{fra87}, where  
part of the stream material was assumed to 
impact the edge of the disk, and the other part was assumed to       
pass above the disk edge to splash further down at smaller radii (the second
hot spot). 
As an example, the Black hole binary Nova Muscae 1991 (LMXB GU Mus, 
observed in quiescence in the optical) reveals the presence of a hot spot 
in its emission-line Doppler maps, consistent with material hitting the disk face around 
phase $\sim 0.6$ near the circularization radius \citep{per15}.  

Algol-type binaries in general display a wider 
and more diverse range of circumstellar structures 
such as e.g. transient disks, gas stream, accretion annulus, and shock regions.
Three-dimensional tomography \citep{aga09}  in short-period Algol-type 
binaries (such as U CrB) detected vertical motion in the region  
where the stream directly strikes the surface of the accreting star,  
while at the (first) hot spot gas is ejected out of the orbital plane 
at velocity $V_z \sim 240-540$km$~$s$^{-1}$. 

For SW Sex stars (a subclass of novalike CVs seen at high inclination), 
a hot spot overflow model was also proposed  \citep*{hel94,hel96}  
to explain the single-peaked Balmer, He\,{\sc i} and He\,{\sc ii} emission 
lines and the strong phase dependence absorption features. 
In these systems the outer radius of the disk is expected to be rather large, 
and the rim of the disk is thickened where the stream hits, 
as well as downstream of that region on the rim of the disk (the elongated
hot spot forms a ``tail''). 
This raised rim region masks the inner disk (around $\Phi=0.8$) and is 
the source of the observed emission lines \citep*{dhi97}. 
Similarly, the famous dwarf nova WZ Sge exhibits a bulge in its 
Doppler map at low accretion rate also supporting a tail 
(with strong H\,{\sc $\alpha$} emission) 
downstream of the hot spot where the material has settled into 
approximate Keplerian motion \citep{spr98}.   
\\

In spite of the complexity of the stream-disk interaction and its many 
different effects on the observations in interacting binaries, we present
here a simple ballistic trajectory approach to describe the phase-dependent 
velocity offsets of absorption lines  in the far-UV (FUV) spectra of accreting
white dwarfs in CVs, which can be applied to and can have implications for 
other systems. 
This model is based on the assumption that in some simple cases the stream-disk
interaction can be described as the supersonic flow (the $L1$ stream) 
past a blunt object (the accretion disk edge) forming a detached bow shock. 
The model is further simplified if the flow is highly supersonic 
(i.e. hypersonic) and nearly isothermal. 
We applied this method to explain the
phase dependence of the absorption-line velocity offsets 
observed in the dwarf novae U Gem in outburst. 
We have chosen U Gem because it has been extensively studied, 
the data are readily available for modeling, and the system does not present
any complex behavior (as do, e.g., RR Pic or IX Vel; see the next section).    
The advantage of this approach 
is that it is much easier and faster to compute than three-dimensional 
hydrodynamic simulations.    

In the next section we present FUV observations of CVs, focusing in particular 
on the phase-dependent absorption-line  velocity offsets observed in U Gem.  
In Sections 3 \& 4 we present the theoretical setup we propose, including
the equations and some tests and preliminary results. In Section 5
we present the results for U Gem, we discuss them in Section 6,
and we close with a summary and conclusion in Section 7.  

\clearpage 

\section{{\bf Stream Overflow Veiling the WD and Inner Disk in CVs}}  

In CVs, the accretion-heated WD is often the dominant 
FUV component at low mass transfer rates,
while at high mass transfer rates the accretion disk becomes  
the dominant source of FUV. 
As a consequence, the main effect of stream overflow in 
FUV spectra of CVs is the
veiling of the hot components, i.e. the WD and/or the inner disk,  
which is observed as a ``dip'' in the FUV light curves of these systems. 
We present here five CV systems (which we have studied in our past FUV spectral
analysis; see Table 1 and references therein) displaying possible signs of 
stream-disk overflow.  

\subsection{{\bf Dips in the FUV Light Curves of CVs}}  

The stream-disk interaction varies from one system to another and 
it also changes as the mass transfer rate increases or decreases within a system.   
Because of that, the UV light curves of CVs exhibit a diverse complexity,  
and we present here a number of systems to illustrate this diversity.  
In Fig.1 we present the FUV light curves of four CVS.  
The light curves were generated by integrating 
the flux, $\int d F_{ \lambda} d \lambda$, 
of individual FUV spectra (exposures) obtained at different orbital phases. 
A summary of the observational details of these archival spectra 
is given in Table 1 together with the inclination of each system.

In the first two panels we show two systems exhibiting a 
dip around phase $\Phi \sim 0.7 \pm 0.1$, where
the dip has an amplitude of $\sim$5\% (IX Vel; top left) and $\sim$15\%
(U Gem; top right).  
IX Vel is  a UX UMa subtype of novalikes 
always found in a high state when the disk dominates the FUV spectrum.   
Even though stream-disk overflow is more pronounced at higher mass transfer 
rate \citep{kun01}, only a  5\% modulation is observed in the FUV 
{\it FUSE} light 
curve of IX Vel, as the system inclination is rather moderate, $57^{\circ}$
(at low inclination, one does not expect any noticeable orbital variability 
due to the stream-disk overflow).    
For U Gem, with an inclination closer to $70^{\circ}$, a stronger
dip is observed at low mass transfer  rate in its {\it Hubble Space Telescope
(HST)} COS light curve, 
more pronounced in the short wavelengths ($\lambda < 1100$\AA ). 
A second dip is also observed around phase 0.25
and was already reported by \citet*{lon06} in the quiescent {\it FUSE} spectrum of U Gem. 
The second dip around phase 0.25 is believed to be due to stream overflow material
bouncing off the disk at phase 0.5 and moving toward phase 0.2-0.3
\citep{kun01}. 
The increase of flux near phase 0.8-0.9 is likely due to the hot spot 
facing the observer. Around phase $\sim$0.0-0.1 
the hot spot is expected to be fully eclipsed \citep{krz65} as the 
disk itself undergoes partial eclipses \citep{sma71,war71,ech07}.  
We have no data with that phase coverage, but the light curve of the
{\it FUSE} data of U Gem in deep quiescence (which has a small gap
between $\Phi\sim 0.98$ and $0.05$) does not show an eclipse. 
Either the eclipse of the hot spot is very narrow, or the disk
might have shrunk such that it is not eclipsed at all. 
Since the outer radius of the disk can increase or decrease
with $\dot{M}$,  
there could be times when the hot spot does not undergo any eclipse.  

In the bottom left panel of Fig.1 we show the {\it HST} STIS FUV light
curve of EM Cyg obtained during two visits (epochs) as the system was 
declining from the peak of an outburst. The first light curve (epoch 
1 in black, as the system is near peak outburst) exhibits a 60\%
drop in flux near phase 0.75, while the second light curve 
(epoch 2 in red, as $\dot{M}$ has decreased one day later) 
has a dip of only 20\%, clearly illustrating how the stream-disk
overflow increases with increasing $\dot{M}$. A 60\% drop in flux 
is rather extreme and unusual, indicating that the density and surface 
area of the obscuring material are large.  

In the bottom right panel, we show the light curve of RR Pic
(an old nova) generated with data covering more than two decades.  
Over that time period, the light curve exhibits a consistent gradual 
drop in flux (reaching 30\%) from phase $\sim 0.0$ to $\sim 0.5$. 
This drop cannot be explained with the {\it usual} stream-disk overflow
masking the inner region of the disk, but rather might be due to a 
more complex configuration as described in \citet{sch03}.  

In the present work, we wish to concentrate on systems displaying a 
modest dip in their FUV light curve around phase $\sim 0.7$ and exclude
systems exhibiting a more complex and/or more extreme behavior.   
In order to further differentiate between the different cases, 
we now look at the spectra obtained at different orbital phases. 

\clearpage 

\begin{figure}[h!]  
\vspace{-6.cm} 
\gridline{
          \fig{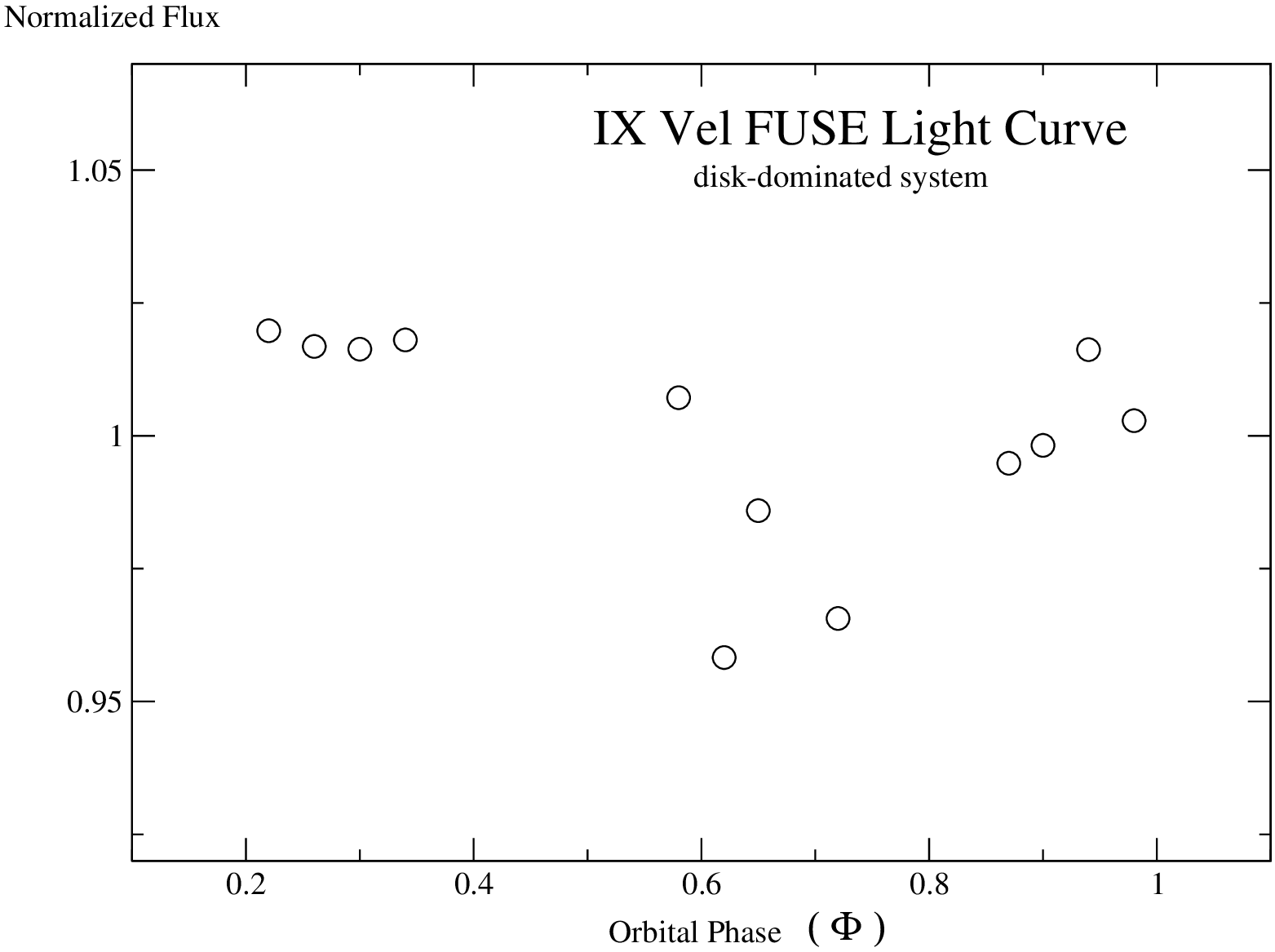}{0.50\textwidth}{}
          \fig{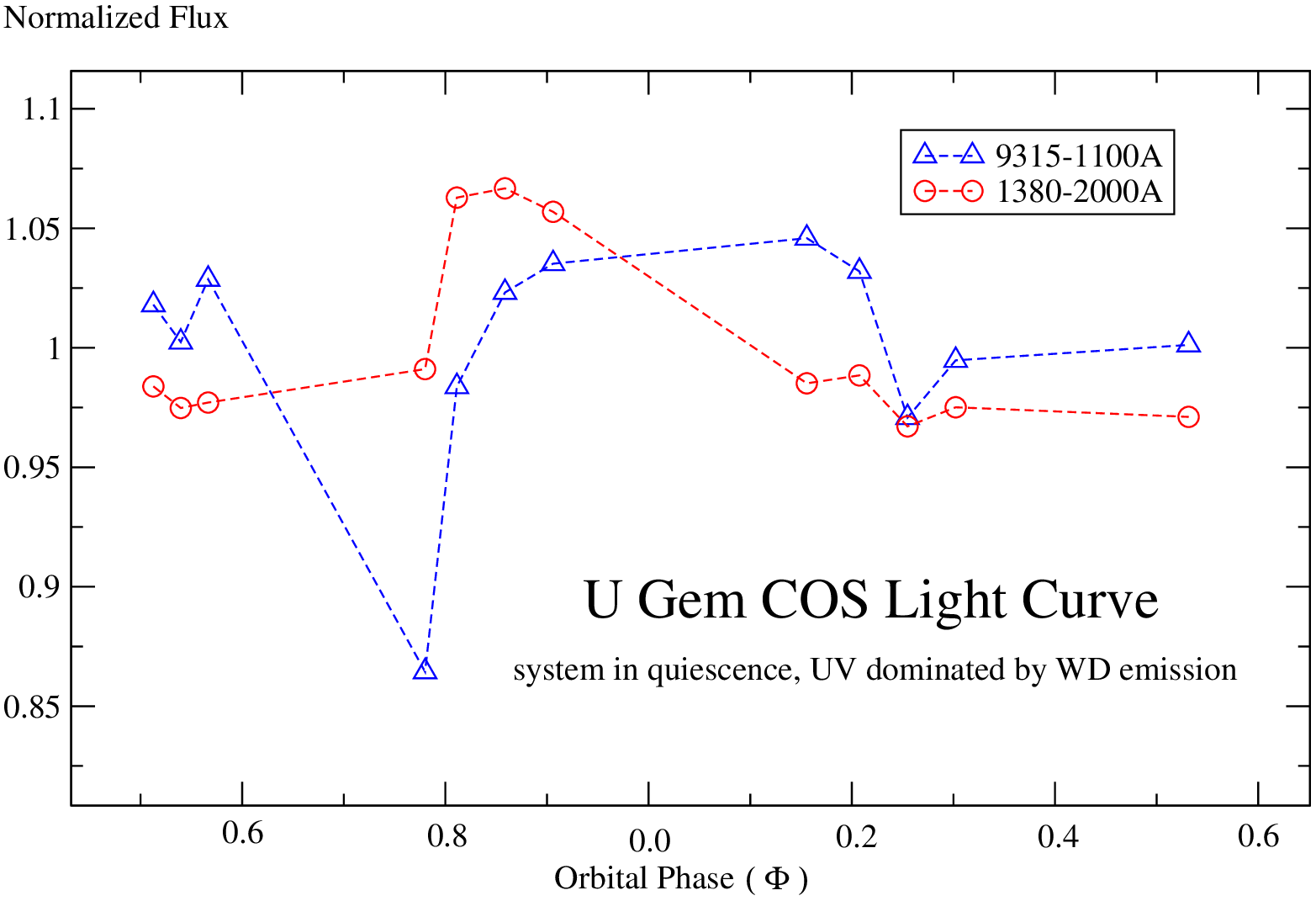}{0.55\textwidth}{}
          }
\vspace{-6.5cm} 
\gridline{
          \fig{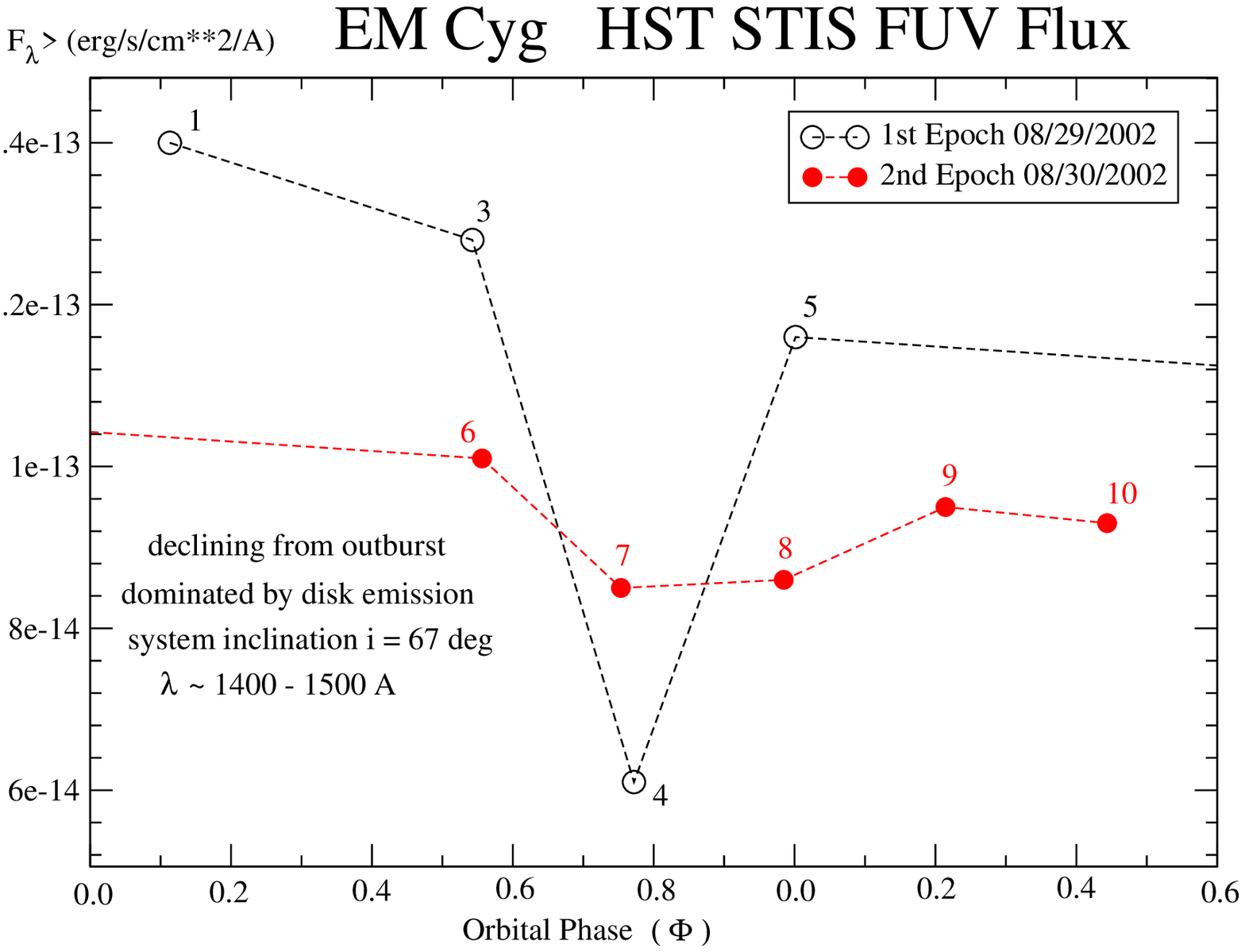}{0.55\textwidth}{}
          \fig{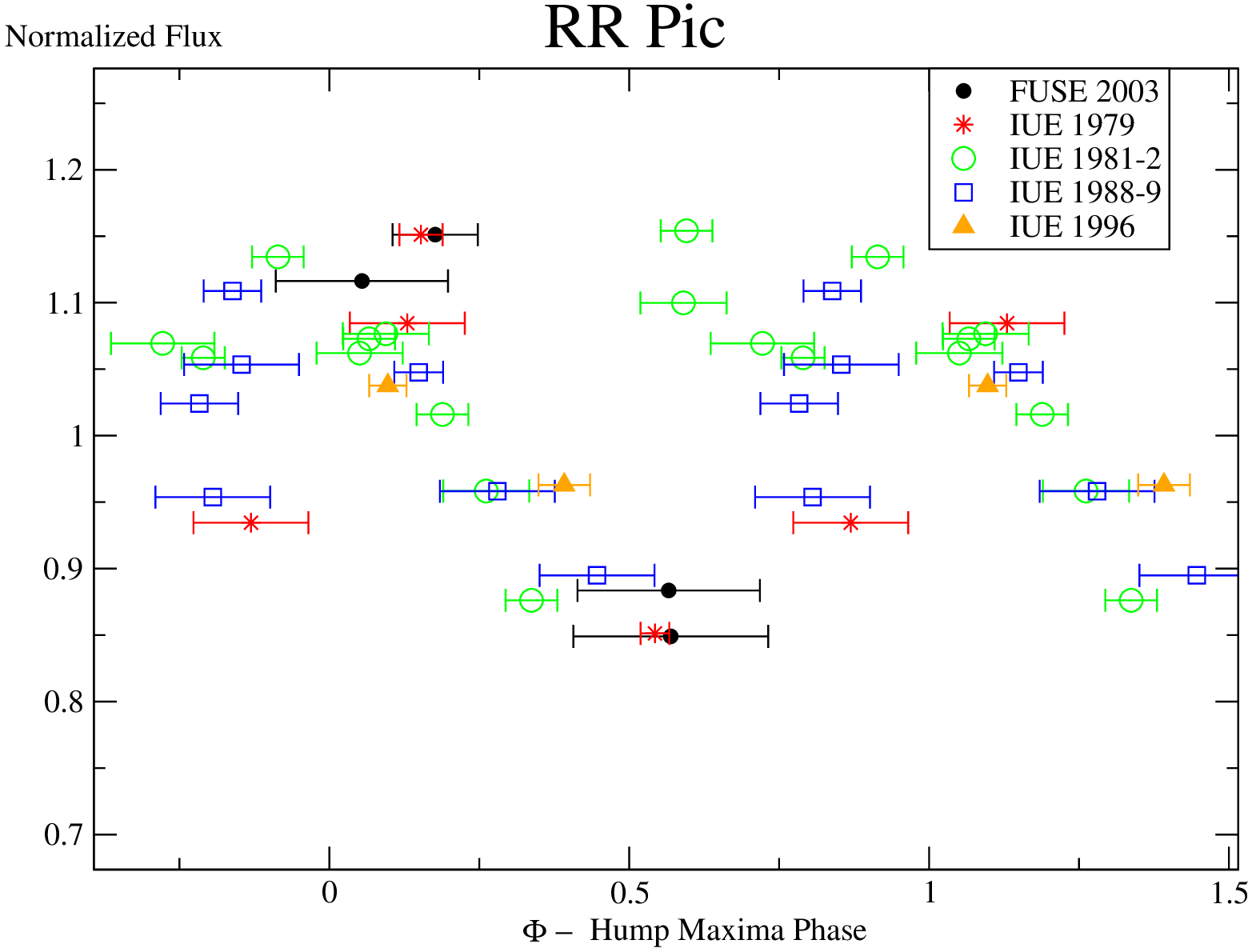}{0.50\textwidth}{}
          }
\caption{
Ultraviolet light curves of cataclysmic variables folded at the 
orbital period. 
{\bf Upper left.} 
The normalized 
{\it FUSE} light curve of the disk-dominated novalike IX Vel;   
it was generated using 12 {\it FUSE} exposures 
and the ephemeris from \citet{beu90}. 
The phase-dependent modulation of the FUV light curve is of the order
of 5\%, a dip is observed around phases 0.6-0.8.    
{\bf Upper Right.} 
The normalized HST/COS light curves
in the wavelength ranges 915-1100 \AA\ (in blue)
and 1380-2000 \AA\ (in red) from the HST COS observations of dwarf nova 
U Gem. The data were
obtained in quiescence, 15 days after the peak of a wide outburst,
when the WD dominates the spectrum.
The phase-dependent modulation of the total FUV flux is of the order of
5\%, but reaches 15\% in the shorter wavelengths.    
(from \citet{god17}). 
{\bf Lower left.}
The HST STIS light curve of the dwarf EM Cyg while the system is 
dominated by emission from the accretion disk. 
Two observations were obtained at one day interval, epochs 1 \& 2 as 
marked, as the system was declining from outburst. 
The dip (around phase 0.7) is larger during the first epoch 
and decreases as the mass transfer rate drops (epoch 2). 
For clarity we did not normalize the flux.  
{\bf Lower right.} 
The normalized (IUE and FUSE) FUV light curve of the nova RR Pic is shown over 
many years. Each year (as indicated in the upper right panel) has been
normalized separately. In spite of the 24 yr coverage, the data show
a consistent drop in flux from orbital phase 0.0 to phase 0.5. The phase shown
is that of the hump maxima \citep{vog17}, the orbital phase itself has
its zero at hump maxima phase $\sim$0.17 \citep{sch03}. 
This light curve agrees very well with the V/R graph of \citet{sch03}.   
} 
\end{figure}

\clearpage

\begin{deluxetable}{lccccccccl}[b!] 
\tablewidth{0pc}
\tablecaption{FUV Spectra of Cataclysmic Variables}         
\tablehead{
System & $i$ & Telescope & Data     & Date       & time     & Exp.time & Sub-Exp. & Figure & reference \\ 
Name   &(deg)&           & Set      & MM-DD-YYYY & hh:mm:ss & (s)      &          & Number &           
}
\startdata 
IX Vel & 57 & FUSE   & Q1120101 & 04-15-2000 & 15:34:47 & 7336  & 12 & 1,2   & \citet{lin07} \\
U Gem  & 67 & COS    & LC1U0101 & 12-20-2012 & 11:54:12 & 5762  & 12 & 1,2,3abc & \citet{god17} \\ 
       &    & "     & LC1U0201 & 12-26-2012 & 04:59:28 &  960  & 4  & 3abc   &  "            \\ 
       &    & "     & LC1U0601 & 01-07-2013 & 23:26:05 &  960  & 4  & 3abc   &  "            \\ 
       &    & "     & LC1U0401 & 01-30-2013 & 22:03:00 &  960  & 4  & 3abc   &  "            \\ 
       &    &FUSE   & A1260112 & 03-05-2000 & 16:36:37 & 2883  & 5  & 3d     & \citet{fro01} \\  
       &    & "     & A1260102 & 03-07-2000 & 10:11:18 & 6248  & 9  & 3d     &  "            \\ 
       &    & "     & A1260103 & 03-09-2000 & 13:50:17 & 7868  & 16 & 3d     &  "            \\ 
RR Pic & 65 & FUSE   & D9131601 & 10-29-2003 & 13:35:41 & 13564 & 4  & 1,2  & \citet{sio17} \\ 
       &    & IUE    & SWP05774 & 07-11-1979 & 21:28:17 &  600  & 1  & 1     & this work  \\  
       &    &  "     & SWP05775 & 07-11-1979 & 23:15:54 & 1200  & 1  & 1     &  "   \\  
       &    &  "     & SWP06624 & 09-24-1979 & 21:50:14 & 2400  & 1  & 1     &  "   \\  
       &    &  "     & SWP06625 & 09-24-1979 & 23:01:54 &  900  & 1  & 1     &  "   \\  
       &    &  "     & SWP15632 & 12-03-1981 & 10:46:47 &  960  & 1  & 1     &  "   \\  
       &    &  "     & SWP15633 & 12-03-1981 & 11:43:02 & 1080  & 1  & 1     &  "   \\  
       &    &  "     & SWP15634 & 12-03-1981 & 12:39:34 & 1080  & 1  & 1     &  "   \\  
       &    &  "     & SWP15635 & 12-03-1981 & 13:33:36 & 1080  & 1  & 1     &  "   \\  
       &    &  "     & SWP15636 & 12-03-1981 & 14:40:08 & 1080  & 1  & 1     &  "   \\  
       &    &  "     & SWP15637 & 12-03-1981 & 15:37:25 & 1080  & 1  & 1     &  "   \\  
       &    &  "     & SWP17748 & 08-23-1982 & 10:39:32 & 1800  & 1  & 1     &  "   \\  
       &    &  "     & SWP17749 & 08-23-1982 & 12:12:42 & 2160  & 1  & 1     &  "   \\  
       &    &  "     & SWP17750 & 08-23-1982 & 13:33:25 & 1800  & 1  & 1     &  "   \\  
       &    &  "     & SWP17751 & 08-23-1982 & 15:17:03 & 1800  & 1  & 1     &  "   \\  
       &    &  "     & SWP17752 & 08-23-1982 & 16:53:07 & 1800  & 1  & 1     &  "   \\  
       &    &  "     & SWP34325 & 09-26-1988 & 16:00:57 & 1620  & 1  & 1     &  "   \\  
       &    &  "     & SWP35522 & 02-10-1989 & 05:28:48 & 2400  & 1  & 1     &  "   \\  
       &    &  "     & SWP35523 & 02-10-1989 & 06:53:51 & 2400  & 1  & 1     &  "   \\  
       &    &  "     & SWP35524 & 02-10-1989 & 08:22:56 & 2400  & 1  & 1     &  "   \\  
       &    &  "     & SWP35525 & 02-10-1989 & 10:12:41 & 2400  & 1  & 1     &  "   \\  
       &    &  "     & SWP36558 & 06-19-1989 & 22:41:13 & 1200  & 1  & 1     &  "   \\  
       &    &  "     & SWP37087 & 09-19-1989 & 22:30:23 & 1020  & 1  & 1     &  "   \\  
       &    &  "     & SWP57074 & 05-08-1996 & 10:41:03 &  780  & 1  & 1     &  "   \\  
       &    &  "     & SWP57075 & 05-08-1996 & 11:39:58 & 1080  & 1  & 1     &  "   \\  
EM Cyg & 67 & FUSE   & C0100101 & 09-05-2002 & 11:34:11 & 8391  &  4 &  2    & \citet{god09} \\  
       &    & STIS   & O6DR0201 & 08-29-2002 & 09:30:33 & 9120  &  4 & 1,2   & this work     \\ 
       &    &   "    & O6DR0202 & 08-30-2002 & 09:33:01 & 11400 &  4 & 1     &  "            \\ 
WZ Sge & 77 & STIS   & O6NF0201 & 07-11-2004 & 07:01:55 & 2300  & 1  & 2     & \citet{god06}    \\ 
       &    &   "    & O6NF0204 & 07-11-2004 & 11:01:31 & 2850  & 1  & 2     &    "             \\ 
\enddata
\tablenotetext{}{
The data presented in this table were post-processed and used in the present 
work to generate  Figures 1, 2, 3, and 4.  Individual exposures 
were extracted and calibrated, and/or presented first
in the references listed in the last column.
Figure 3d was taken directly from \citet{fro01}.
\vspace{2.cm}  
} 
\end{deluxetable} 

\clearpage 

\subsection{{\bf The Dip Spectra}} 

We now compare the ``dip'' spectrum (obtained near $\Phi \sim 0.7$) to the 
(assumed) unaffected
spectrum for the 5 CV systems listed in Table 1.   
The systems exhibiting a dip of the order of 5-10\% (Fig.1) show deeper absorption lines
and the appearance of new lines around phase $0.7$ (e.g. IX Vel and U Gem,
Fig.2, upper panels). 
Namely, the light curve dip near phase $\Phi \sim 0.6-0.8$ is 
due mainly to the presence of deep absorption lines \citep{fro01,god17}. 
Occasionally the small amplitude of the dip near phase 0.7 is due to a small drop
in the continuum flux level as shown for the quiescent spectra of 
WZ Sge and EM Cyg in Fig.2 (middle panels).  
As expected, the flux drop in the FUV spectra of accreting white
dwarfs is even more dramatic in system accreting at a high rate.   
In the lower panels of Fig.2 we present the disk-dominated (high
transfer rate) spectra of EM Cyg (in outburst) and RR Pic
and how they are affected by the stream-disk overflow. 
The drop in the continuum flux level is very large; the stream material
must be dense and have a large projected 
surface area to obscure a large fraction of
the inner disk.   

The degree by which the FUV spectrum is attenuated (appearance of absorption
lines and/or drop in the continuum flux level) clearly depends on the density 
and temperature of the stream material.  
As shown by \citet{lon06}, using synthetic stellar spectra 
to model the FUV spectra of U Gem, the veiling stream material    
has a temperature of 10,000-11,000~K with
a density of $10^{13}$~cm$^{-3}$ at phase $\sim 0.7$.   

The {\it contamination} of the FUV spectra of CVs with 
absorbing stream material requires that any spectroscopic
analysis be carried out only at the unaffected orbital phases 
as was done, e.g. by \citet{lon06,god09,sio17}. However, one cannot
rule out that veiling might actually occur at all orbital phases 
\citep{lon06} and at all inclinations,
and it is, {\it a priori}, not clear which absorption line  forms 
in the white dwarf photosphere and/or disk, and which absorption line forms
in the veiling material.    
One way to try and identify the location of the formation of the absorption lines  
is to look at their velocity shift, as shown in the next subsection.

\clearpage

\begin{figure}
\vspace{-6.cm} 
\gridline{
          \fig{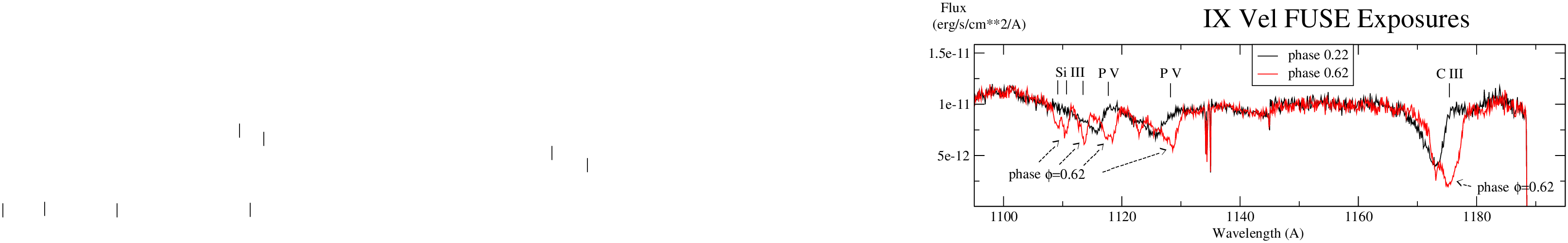}{0.45\textwidth}{}
          \fig{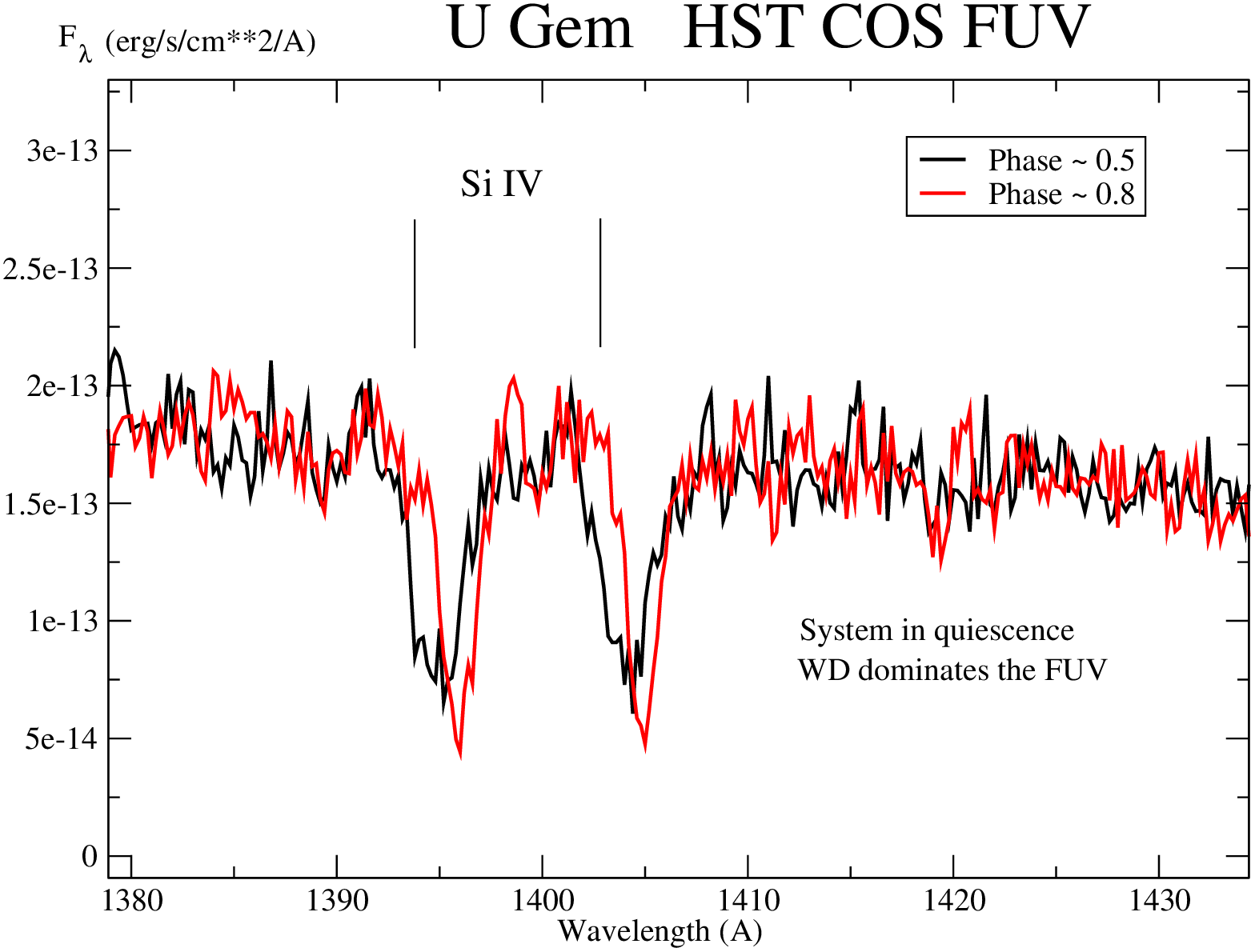}{0.40\textwidth}{}
          }
\vspace{-3.5cm} 
\gridline{
          \fig{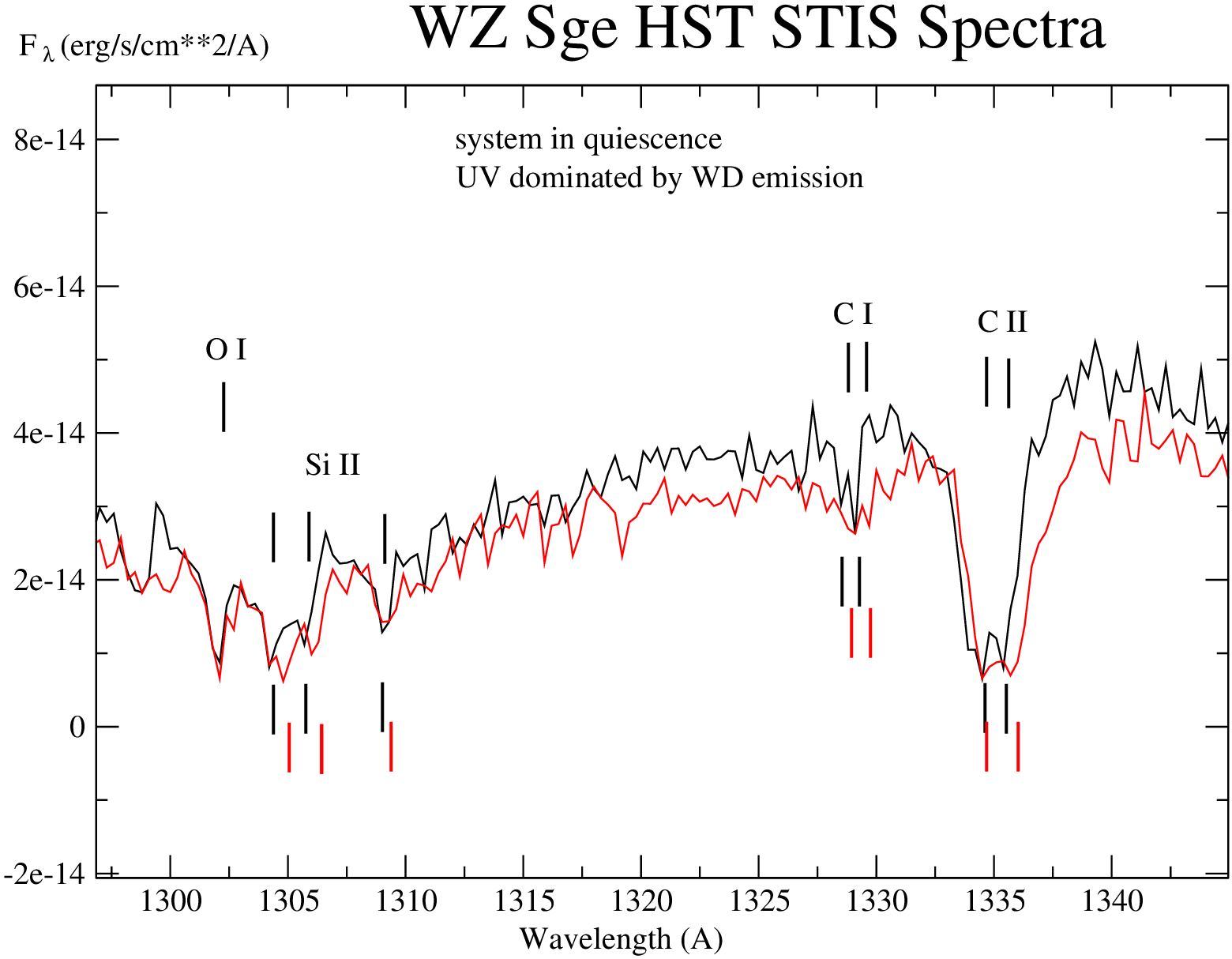}{0.40\textwidth}{}
          \fig{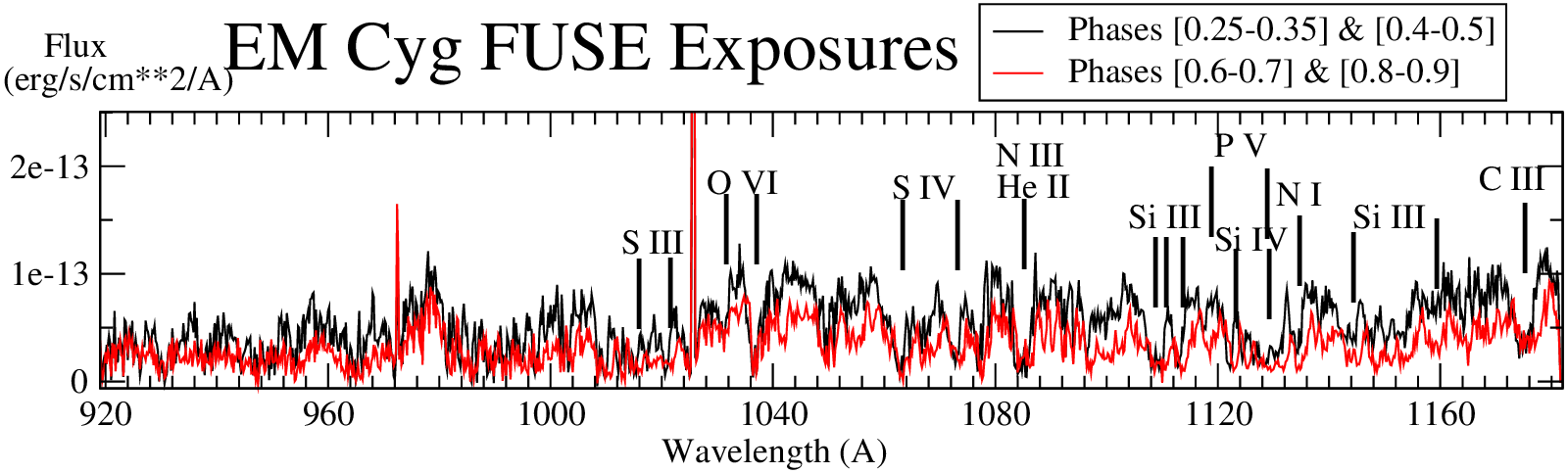}{0.45\textwidth}{}
          }
\vspace{-5.0cm} 
\gridline{
          \fig{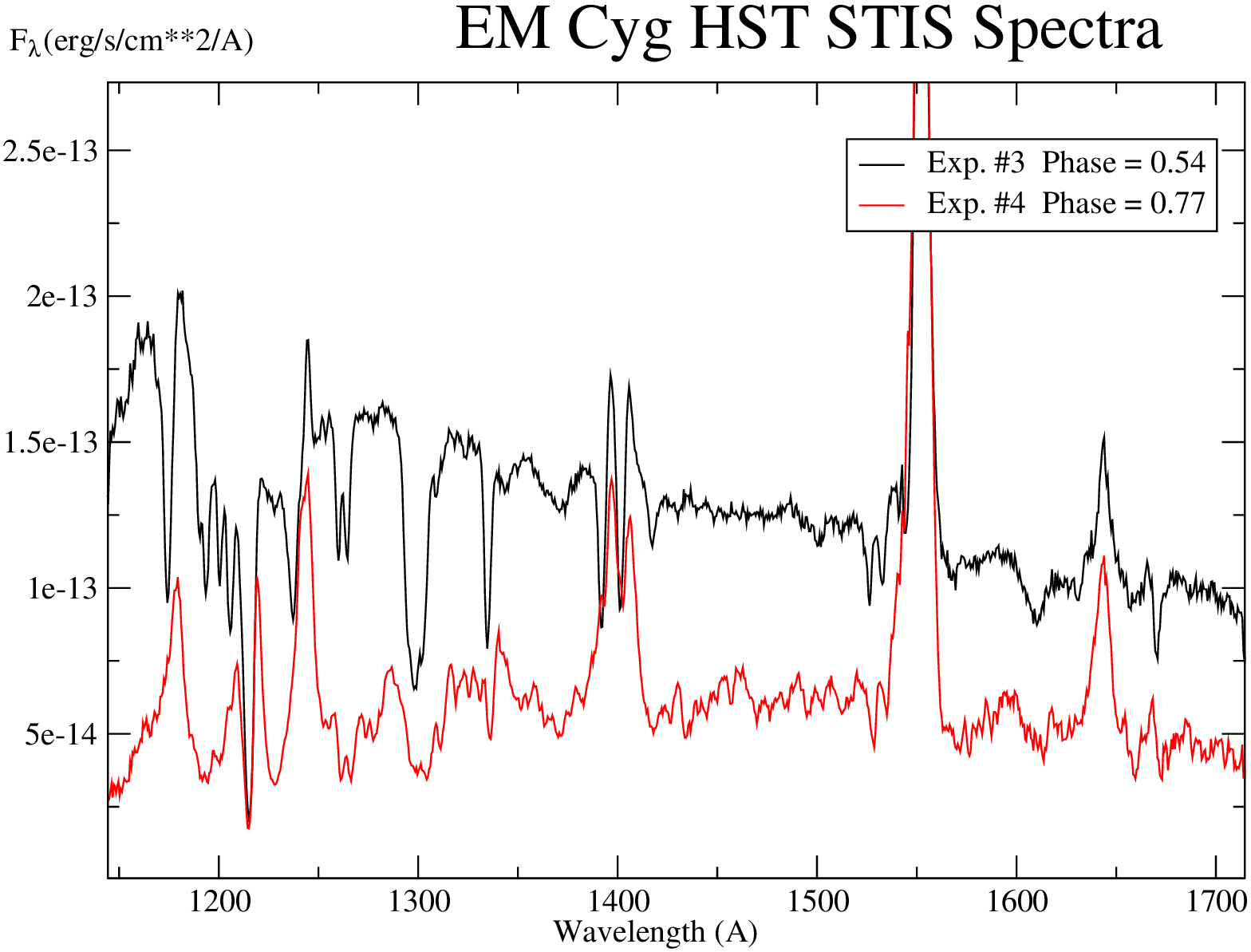}{0.45\textwidth}{}
          \fig{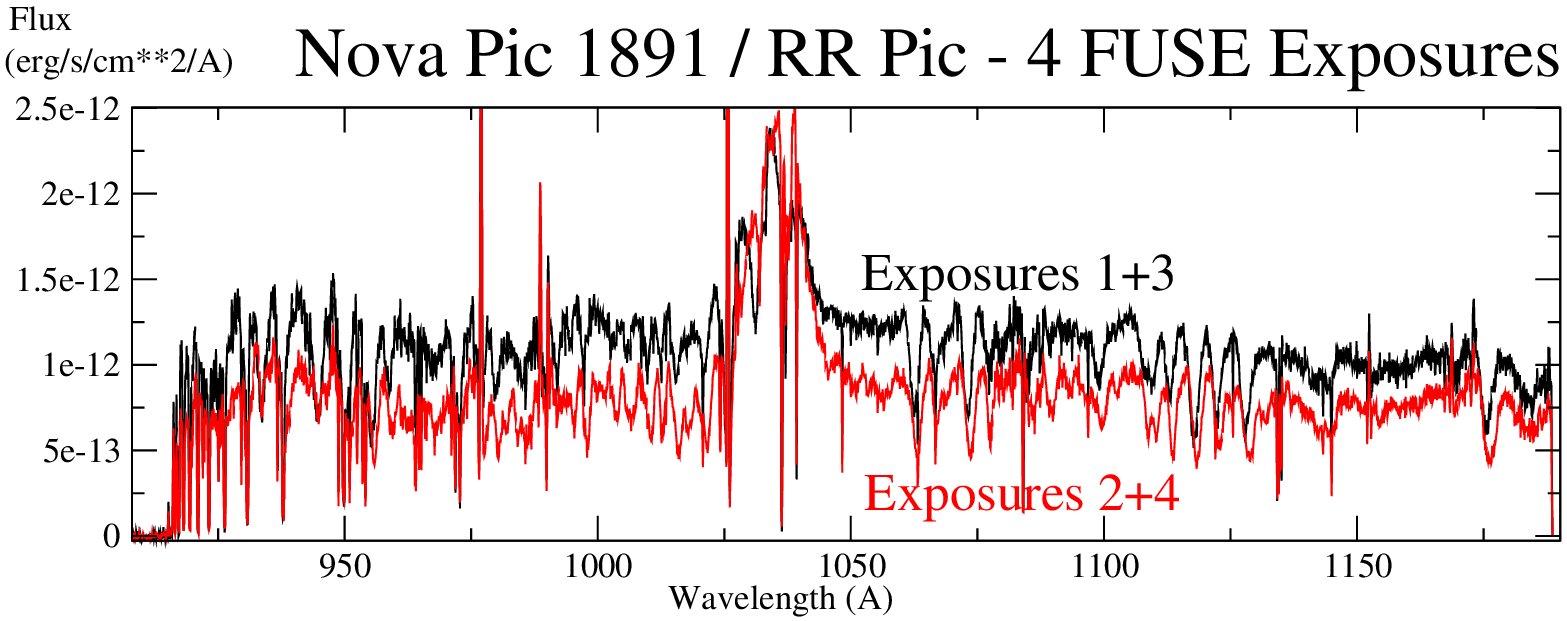}{0.45\textwidth}{}
          }
\vspace{-2.0cm} 
\caption{{\bf Upper Left.}   
Two of the {\it FUSE} exposures that were used to generate the light curve
of IX Vel (disk-dominated) in Fig.1 are shown here at orbital phases 
$\Phi=0.62$ (in red) and $\Phi=0.22$ (in black).  The 5\% drop in flux observed  
near phase $\Phi \sim 0.6-0.8$ is due mainly to the appearance 
of deep absorption lines as the stream material slightly veils 
the inner parts of the accretion disk. 
{\bf Upper Right.} 
Similarly, U Gem observed in quiescence reveals deeper absorption lines
around phase $\sim 0.8$ (in red) compared to phase 0.5 (in black), 
while the continuum flux level remains the same.  
{\bf Center Left.} 
The HST STIS spectra of WZ Sge in quiescence exhibits a slight decrease
in flux around phase 0.75 (in red)  compared to other orbital phase (in black).  
{\bf Center Right.}  
The {\it FUSE} spectrum of the dwarf nova EM Cyg in quiescence  
exhibits a significant drop in the continuum flux level around phase
$\Phi\approx 0.6-0.9$ (in red) due to stream material veiling the white dwarf. 
Even though the mass transfer rate is low, the stream overflowing the disk
edge is capable to significantly reducing the FUV emission from the WD. 
{\bf Lower Left.} 
The HST STIS spectra (exposures 3 and 4 in Fig.1) of EM Cyg in outburst   
reveal a strong veiling of the inner disk, reaching 60\% of the continuum
flux level around phase 0.77 when compared to phase 0.54.  
{\bf Lower Right.}  
RR Pic FUSE spectra obtained around phase 0.5 (in red) exhibit a 20\% drop
in flux compared to the spectra obtained near phase 0.0 (in black). The
gradual decrease from phase 0.0 to 0.5 cannot be explained with the 
usual stream-disk overflow material veiling the inner disk, as        
the system has a rather unusual symmetry \citep{sch03}.  
}
\end{figure}

\clearpage

\subsection{{\bf Orbital Variability of the Absorption-Line Velocity} } 

One of the CV systems studied extensively for which
a large amount of FUV data is available is the prototypical dwarf
nova U Gem. Since the system has been analyzed previously, we use 
the results of these analysis here to illustrate the orbital
variability of the absorption-line centers. Some of the lines
are thicker than others, while some lines appear only around phase
$\sim 0.7$ and $\sim 0.2$ \citep{fro01,lon06,god17}.     
In Fig.3 the radial velocity shift (in km/s) is drawn against the binary
orbital phase $\Phi$ for each absorption line when the system was
in quiescence \citep{god17} and in outburst \citep{fro01} 
(the observation log for the data is presented in Table 1).  

The solid line represents the WD velocity
shift (including the recessional system velocity). 
In the first three panels (a, b, and c), U Gem was observed with 
{\it HST}/COS  in quiescence. 
Expecting the lines to form in the WD photosphere, the
WD gravitational redshift has been included to the WD velocity shift. 
Each species is shown with a different symbols in each panel as indicated
in the upper left.  
Panels (a) and (b) show that the carbon lines and the long-wavelength 
silicon lines follow more closely the WD than the 
shorter-wavelength silicon lines in panel (c).  
The higher velocity (red-) shift of the lines in panel (c) indicates that
the short-wavelength silicon lines form in material that is falling 
toward the WD at most phases.   
The carbon lines follow the WD more closely than the other lines,
an indication that the carbon lines are probably forming in the 
WD photosphere. 

In the last panel in Fig.3 (d, lower right), we display U Gem velocity offsets 
from rest vs. orbital phase for the centers
of several absorption lines for three successive {\it FUSE} observations during
an outburst (the panel was taken from \citet{fro01}).   
The solid line represents the orbital motion of the WD without the gravitational
redshift since the lines are not expected to form in the WD photosphere.
The most striking feature is the much higher velocity offsets between
phases 0.6 and 0.8 for the open and filled circles, which are attributed
\citep{fro01} to the inflowing stream overflow material moving into the 
line of sight, which is not clearly observed in quiescence.
In quiescence (panel c), only one line shows such a departure at only
one phase (Si\,{\sc iv} 1128.3 at phase 0.78 with $V_R \approx 620$km/s).  

\cite{fro01} also remarked that the lines are blue shifted around phase 0.2-0.4 
(Fig.3d) and might indicative of the velocity of the material 
bouncing off near phase 0.5 toward phase 0.3, the velocity  
offsets near phase 0.3 are rather scattered near the WD velocity similar 
to what is observed in quiescence (panels a, b, and c) at all orbital phases.
We will come back to this in Section 6.

\clearpage

\begin{figure}
\vspace{-2.5cm} 
\gridline{\fig{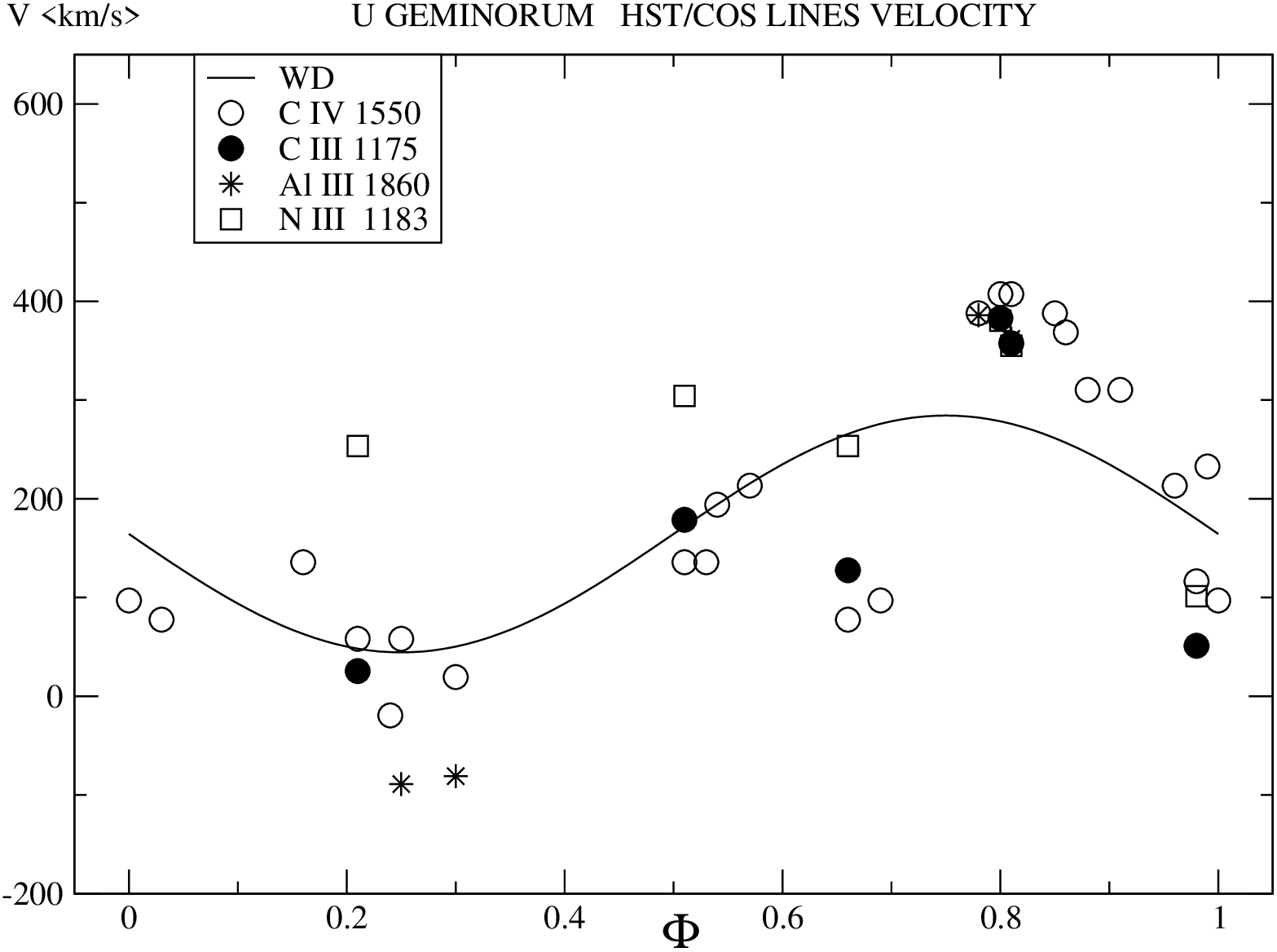}{0.44\textwidth}{\vspace{-0.5cm}(a)}
          \fig{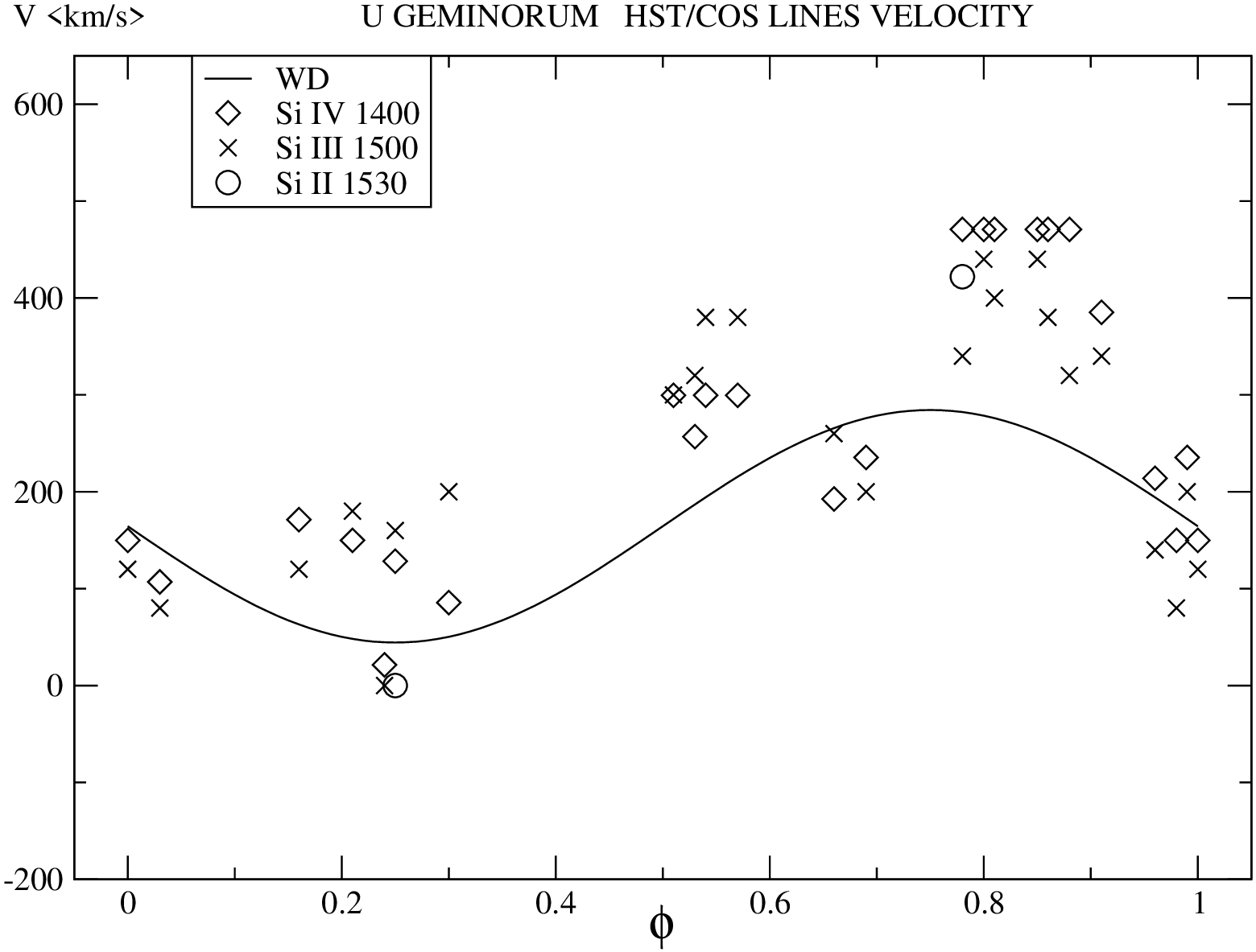}{0.44\textwidth}{\vspace{-0.5cm}(b)}
          }
\vspace{-1.5cm} 
\gridline{
          \fig{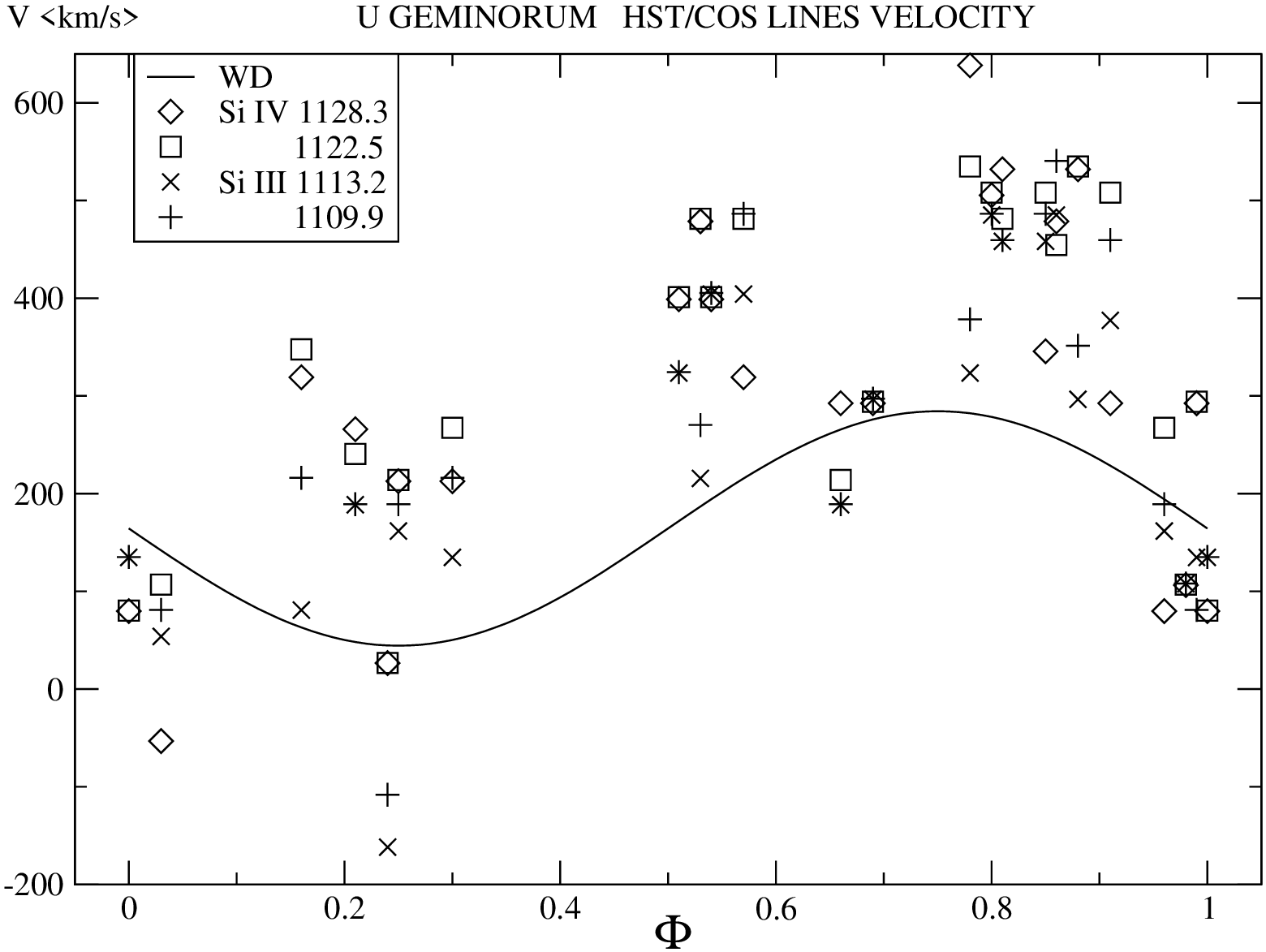}{0.44\textwidth}{\vspace{-8.cm}(c)}
\hspace{-1.cm}  \fig{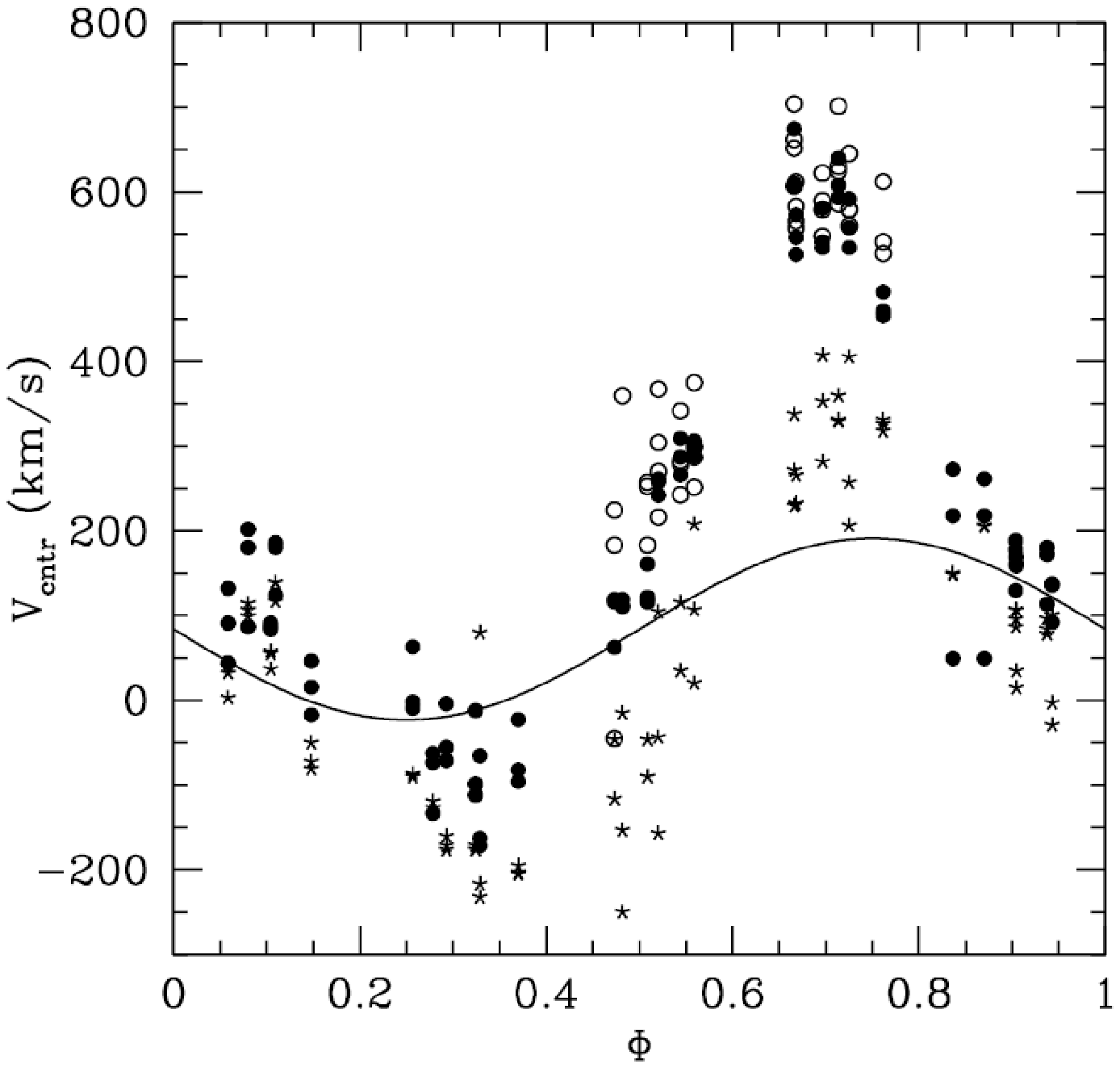}{0.69\textwidth}{\vspace{-8.cm}(d)} 
          }
\vspace{-8.0cm} 
\caption{
U Gem FUV absorption lines velocity shift (in km/s) as a function of the 
binary orbital phase $\Phi$. The solid line represents the WD velocity
shift.
In the first three panels (a, b, and c), U Gem was observed 
with HST/{\it COS} in quiescence when the 
WD is expected to dominate the FUV. 
The species are as indicated in the panels. 
(the data is taken from \citet{god17}). 
In the last panel (d, lower right) we display U Gem velocity offsets 
from rest vs. orbital phase for the centers
of several absorption lines for 3 successive {\it FUSE} observations during
an outburst (the figure is taken from \citet{fro01}).   
Filled circles are for 
S\,{\sc iv} (1073), P\,{\sc v} (1118), Si\,{\sc iv} (1122); 
open circles are for N\,{\sc iii} (980),
S\,{\sc iii} (1021), C\,{\sc iii} (1175). 
The asterisks are for highly ionized species
S\,{\sc vi} (944) and O\,{\sc vi} (1032, 1038). 
The velocity offsets between phases 0.6 and 0.8 for the open and filled circles
is much larger than at other phases, due to the 
in-flowing stream material moving into the line of sight. 
}
\end{figure}

\clearpage 

As another example of absorption-line velocity offsets,
we chose IX Vel because it is one of the few other systems with a large 
number of successive {\it FUSE} spectra  covering almost the
entire orbital phase.  We extracted the 
velocity offsets of its most prominent and reliable absorption lines
from the 12 exposures (used to generate its FUV light curve in Fig.1a).  
These lines are the S\,{\sc iv} (1063 \& 1073) lines,  
P\,{\sc v} (1118 \& 1128) lines, and the C\,{\sc iii} (1175) multiplet. 
We find that most of the lines are blue shifted by up to $\sim 600$km/s
which is due to its strong disk wind.  
Indeed, IX Vel exhibits a variable wind that is responsible for narrow blue shifted
absorption lines (on top of broad absorption lines) with velocities of -900km/s
\citep[as analyzed in its HST spectrum by][]{har02}. A similar sharp
absorption feature
can be seen in the C\,{\sc iii} (1175.3) absorption line in Fig.2 but
with a smaller velocity of $\sim -500$km/s, while the H\,{\sc i} (1025.7,
not shown) line has a wind component velocity of -670km/s. 

We draw IX Vel FUV line velocity offsets against the orbital
phase in Fig.4. The overall pattern is similar to that of U Gem (Fig.3d) in  
that near phase
$\sim 0.6-0.7$ the line velocity offsets are markedly displaced {\it upward}. 
The main difference here is that {\it all} the lines seem to have been
blue shifted by about -450km/s. One way to interpret this is that   
the wind component is literally ``blowing away'' the absorbing 
gas (above the disk and in the stream) towards the observer at all phases. 
By blue shifting the WD radial velocity by
450km/s (dashed line in the figure), we obtain a figure similar to 3d.   
However, the disk wind is deflecting the stream material 
too, affecting its trajectory; consequently, the pattern 
obtained near phase 0.6-0.7 (Fig.4) might have been shifted left or right, 
as well as up or down, due to the disk wind.  

The line velocity offsets provide important dynamical information 
on the stream overflow material  as it is veiling the inner region 
of the disk and WD near phase 0.7.   
While Fig.4 is instructive, we cannot use these data to model the stream
overflow, as the wind would have to be taken into account.  
Instead, in this work, we use the line velocity offsets of
U Gem to help model the stream overflow. Our aim is to model the
stream overflow material as it veils the inner disk and obtain 
radial velocities matching the velocity offsets of the lines 
in the {\it FUSE} spectrum of U Gem in outburst (Fig.3d).

\begin{figure}[b!]   
\vspace{-12.cm} 
\plotone{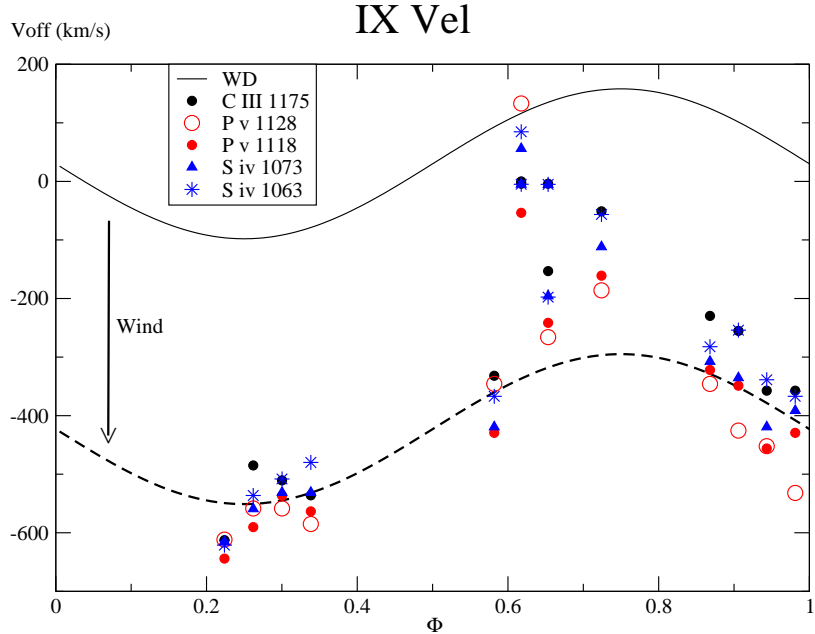}  
\caption{
Velocity offsets (in km/s) of the most prominent absorption lines in the 
{\it FUSE} spectrum of IX Vel as a function of the orbital phase $\Phi$. 
IX Vel is always in state of high accretion rate where the disk 
dominates the FUV. The lines have been marked 
with symbols as indicated in the upper left. A WD velocity amplitude 
of 138km/s was taken from \citet{beu90} (solid black line).  
The lines are strongly blue shifted at most orbital phases. 
To account for the excess negative velocities, we assume a -450km/s 
wind velocity, and we move the WD radial velocity 
down by 450km/s (dashed black line). In such a configuration (dashed line)
the overall pattern of the line velocity offsets as a function of the 
orbital phase is similar to that of U Gem (Fig.3d).  
}
\end{figure}

\clearpage

\section{{\bf The Ballistic Trajectories}} 

Our goal is to {\it numerically} 
follow the $L1$ stream as it leaves the $L1$ region,
impacts the disk edge where it is deflected above the disk, 
and eventually lands onto the disk near phase 0.5-0.6.
The radial velocity of the deflected stream in the line of sight of
the observer is then to be compared to the radial velocity shift
of the absorption lines in the {\it FUSE} spectrum of U Gem in outburst.  
In this section we present the numerical solver we have developed to 
carry out this task, under the assumption that the $L1$ stream can be 
represented by test particles following ballistic trajectories within 
the context of the Roche model in the restricted three-body problem.  
The specific description of the deflection of the stream by the disk edge
is delayed to the next section.  

To avoid any confusion, we stress here that 
in the manuscript we use the following notation (see also Figs.5, 6, \& 7): 
$i$ is the inclination of the binary plane;
$\Phi$ is the binary orbital phase; 
$\theta_s$ is the angle the $L1$ stream is making with the line joining
the two masses as the stream leaves $L1$; 
$\varphi_0$ is the angle the $L1$ stream is making  with the X-direction at the
point it impacts the outer edge of the disk (before it is deflected);  
$\phi$ is the impact angle, it 
is measured counterclockwise from the line joining 
the two stars to the line joining the primary star to the 
site of impact of the stream on the disk; 
$\theta$ is the angle the bow shock makes with the horizontal (at any
point $P$ on its surface), which
under certain circumstances is equal to the 
launching angle at which the test particles are 
ejected off the shock front
(the launching angle has a maximum $\theta_{\rm max}$);   
$\varpi_{\rm d}$ is the circularization radius, namely  
the radius at which the angular velocity of the test particle equals
the Keplerian velocity; 
and $\varphi_{\rm h}-180^{\circ}$ is the circularization angle, 
the angle at which the stream impacts a disk of radius 
$R_{\rm disk} = R_{\rm d} = \varpi_{\rm d}$, measured counterclockwise
from the line joining the two stars; 
in that case $\phi = \varphi_{\rm h} - 180^{\circ}$
\citep{lub75}.

\subsection{{\bf Assumption and Equations}}

In the vicinity of the first Lagrangian point $L1$, 
the hydrostatic balance becomes less well
satisfied as the stream flows, and the pressure force expands the stream
width and thickness, increasing the stream scale height, until vertical
forces drive the stream back toward the disk midplane. As the flow becomes
hypersonic near the disk edge, the velocity field can then be regarded as 
nearly horizontal and  
the center of the gas stream can be described with test particles
following ballistic trajectories in the restricted three-body problem  
\citep[][as long as the ballistic trajectories/orbits do not intersect]  
{lub75,lub89}.

The gaseous $L1$ stream is modeled by a finite number ($N=91$) of 
test particles moving under the influence of the gravitational potential
of the binary masses.   
For each test particle, with coordinates $(x,y,z)$, 
we write the equations 
within the circular restricted three-body problem in the rotating 
(center of mass) frame of reference \citep{mou14,plu18}  
\begin{large} 
\begin{equation}
\left\{ 
\begin{array}{l} 
\frac{dx^2}{dt^2} =  +2\frac{dy}{dt}n + x n^2  
-Gm_1\frac{(x-x_1)}{r_1^3} -Gm_2\frac{(x-x_2)}{r_2^3} ,  
\\ 
\frac{dy^2}{dt^2} =  -2\frac{dx}{dt}n + y n^2  
-Gm_1\frac{y}{r_1^3} -Gm_2\frac{y}{r_2^3} ,  
\\ 
\frac{dz^2}{dt^2} =  
-Gm_1\frac{z}{r_1^3} -Gm_2\frac{z}{r_2^3} ,  
\end{array}
\right. 
\end{equation}
\end{large} 
$$ 
{\rm where } ~~~r_1=\sqrt{ (x-x_1)^2 + y^2 + z^2 }, ~~~~~~~~
r_2=\sqrt{ (x-x_2)^2 + y^2 + z^2 },
$$ 
where $G$ is the gravitational constant, $m_1$ is the mass of the WD
(located at $(x_1,0,0)$), $m_2$ is the mass of the secondary 
(located at $(x_2,0,0)$), $n=\sqrt{G(m_1+m_2)/a^3}$ and
$a$ is the binary separation. 
We chose the units of length, mass and time such that (respectively)  
$a=1$, $m_1+m_2=1$ and $G=1$, which gives $n=1$.  
The equations simplify to 
\begin{large} 
\begin{equation}
\left\{ 
\begin{array}{l} 
\frac{dx^2}{dt^2} =  +2\frac{dy}{dt} + x   
-(1-\mu)\frac{(x-x_1)}{r_1^3} -\mu \frac{(x-x_2)}{r_2^3} ,  
\\ 
\frac{dy^2}{dt^2} =  -2\frac{dx}{dt} + y   
-(1-\mu)\frac{y}{r_1^3} -\mu \frac{y}{r_2^3} ,  
\\ 
\frac{dz^2}{dt^2} =  
-(1-\mu) \frac{z}{r_1^3} -\mu \frac{z}{r_2^3} ,  
\end{array}
\right.   
\end{equation}
\end{large} 
where we put $m_1= 1-\mu $ and $m_2=\mu$
(with $\mu < 1/2$). 

\clearpage 

\subsection{{\bf The Second-order Runge-Kutta Method}}

The three second-order equations (2)  
are written as six first-order equations 
\begin{large} 
\begin{equation} 
\left\{ 
\begin{array}{l} 
\frac{dv_x}{dt} = 2 v_y + \frac{\partial F}{\partial x} , 
\\           
\frac{dv_y}{dt} = -2 v_x + \frac{\partial F}{\partial y} , 
\\           
\frac{dv_x}{dt} = \frac{\partial F}{\partial z} , 
\end{array}
\right. 
~~~~~~~~~~~~~~~~~~~~~ 
\left\{ 
\begin{array}{l} 
\frac{dx}{dt} =  v_x , 
\\      
\frac{dy}{dt} =  v_y , 
\\      
\frac{dz}{dt} =  v_z , 
\end{array}    
\right.  
\end{equation}
\end{large} 
with 
$$ 
F =  \frac{1}{2} \left( x^2 + y^2 \right) + 
\frac{1-\mu}{r_1} + \frac{ \mu}{r_2} .
$$ 
Or in a more compact notation, we have  
$$
\frac{df_i}{dt} = {\cal{F}}_i,~~~~~~~~(i=1,2,..6),  
$$
where ${\cal{F}}_i$ is a function of $f_j$'s $(j=1,2,..6)$.

We use a second-order Runge-Kutta method 
(``predictor-corrector''; e.g. \citet{numrec}) to numerically integrate the
particle trajectory and advance from $f_i^0$ to $f_i^1$ in a time
$\Delta t$:  
$$ 
\left\{ 
\begin{array}{l} 
f_i^{1/2} = f_i^0 + {\cal{F}}_i^0 \frac{\Delta t}{2},   
\\ 
f_i^1     = f_i^0 + {\cal{F}}_i^{1/2} \Delta t, 
\end{array}
~~~~~~~(i=1,2,..6).  
\right. 
$$ 
The numerical integration of the equations to follow  
the ballistic trajectories of the 
test particles is begun in the vicinity of the
first Lagrangian point $L1: (x_0,y_0,z_0)$.
This completes the description of our restricted three-body numerical solver.  

\subsection{{\bf The First Lagrangian Point - L1}}  

The first Lagrangian point is an extremum of the
Jacobi integral ($F$ + constant) 
located on the X-axis, and it is found by solving 
the quintic equation derived from the condition 
$\partial F/\partial x=0$. 
The quintic equation has the form 
\begin{equation}
\rho^5 - (3-\mu) \rho^4 + (3-2\mu) \rho^3 
-\mu \rho^2 +2\mu \rho -\mu =0 ,
\end{equation} 
where $\rho=x_2-x$, $x-x_1=1-\rho$, the first mass
$1- \mu$ is located at $x_1$ and the second mass $\mu$
is located at $x_2$. 
The real positive root of this equation is given by 
\citep{mou14,plu18} 
\begin{equation} 
\rho = 
\left( \frac{\mu}{3} \right)^{1/3} - 
\frac{1}{3} \left( \frac{\mu}{3} \right)^{2/3} -  
\frac{1}{9} \left( \frac{\mu}{3} \right)^{3/3} -
\frac{23}{81} \left( \frac{\mu}{3} \right)^{4/3} + ... 
\end{equation} 
In the present work, we find the root of the equation by numerically 
increasing $\rho$ from 0 to 1, in steps of $10^{-6}$, namely, the 
$x$-coordinate of the first Lagrangian point, $x_{l1}$ is accurate
to the sixth digit.

\subsection{{\bf The Roche Lobe Radius}}  

A convenient measure of the size of the Roche Lobe, and therefore
of the maximum size of the accretion disk, is given by the
radius $R_L$ of a sphere which has the same volume as the
Roche lobe \citep{kop59}. This radius can be approximated by
the following expressions \citep{pac71} 
\begin{large} 
\begin{equation}
\left\{ 
\begin{array}{lr}  
\frac{R_L}{a} = 0.38 + 0.2 \log{ \frac{m_1}{m_2} } , 
~~~~~ & 0.523 \le \frac{m_1}{m_2}   \le  20  , 
\\ 
\frac{R_L}{a} = 0.46224 \left( \frac{m_1}{m_1+m_2} \right)^{1/3}, 
~~~~~ &    0 \le \frac{m_1}{m_2}   \le  0.523 . 
\end{array} 
\right.  
\end{equation} 
\end{large}

\subsection{{\bf Test Without and With a Disk} }

Since the $L1$ point is an equilibrium point, in order to integrate the
ballistic trajectory of a particle, we need to place the particle 
at an infinitesimal distance $\epsilon > 0$ from $L1$ and 
give it a small velocity $\eta >0$, 
otherwise the CPU time it takes for the particle to move is very long. 
This also ensures that the particle starts moving in the desired direction
(i.e. in the Roche lobe of the primary star). 
We denote the coordinates of $L1$ by $(x_{L1},0,0)$, and place the 
particle on the X-axis at $x_{L1} + \epsilon$ with a velocity $v_x= \eta$. 
We first chose $\epsilon$ and $\eta$ large such that 
a test particle leaves the $L1$ region in a finite number 
of time steps (i.e. a few thousands). Next, we reduce the values 
$\epsilon$ and $\eta$ until we observe that the integrated trajectory remains 
constant, which gives $\epsilon = 10^{-6}$ and $\eta = 10^{-3}$.
The time step is initially chosen to be very small, then it is increased
until it is found to affect the accuracy of the integration. 
We find that $\Delta t = 10^{-4}$ is large enough to carry out the integration
in a reasonable CPU time, while it is small enough to give a good        
accuracy (as described below). 

As an actual test of the ballistic trajectory solver, we chose to replicate 
some of the results obtained by \citet{lub75}. We follow the ballistic 
trajectory of a test particle for different values of the 
binary mass ratio, {\it first without a disk}, 
and compute several parameters for comparison
with the results of \citet{lub75}. 
The first  parameter is the angle $\theta_s$ that the gas stream is making
with the X-axis as it leaves the $L1$ region. Since we are integrating
the equations from $x_{L1}$, we measure $\theta_s$ by solving for 
$a$ and $b$ as
$$
\theta_s = \lim_{x \rightarrow x_{L1}} (x - x_{L1} ) a + b . 
$$  
The next parameter is the circularization radius $\varpi_{\rm d}$,
i.e. the radius at which the angular velocity of the test particle equals
the Keplerian velocity. 
We then place a disk of radius $R_{\rm disk}= \varpi_{\rm d}$, and 
measure the circularization angle $\varphi_{\rm h}-180^{\circ}$, 
the angle $\phi$ at which the $L1$ stream impacts the disk edge, measured
counterclockwise from the line joining the two masses.  

The values we obtained are listed in Table 2. We checked 
these values against those of \citet{lub75} and found them exact to 
within up to a few $10^{-4}$ of their relative values.
This is about $0.01^{\circ}$ in the impact angle $\phi$, which is 
much smaller than the width of the phase angle $\Phi$ at which 
line velocity offsets are obtained, since the data is often binned to 
more than several 100s, corresponding to a few degrees
($\sim 1/100$ of the orbital periods).  

\begin{deluxetable}{cccccc}
\tablewidth{0pc}
\tablecaption{Test of the Ballistic Trajectories}
\tablehead{
Mass Ratio & $\theta_s^{(1)}$ & $\theta_s^{(1)}$ & $x_{L1}^{(2)}$ & $\varpi_{\rm d}^{(3)}$ & $\varphi_{\rm h}-180^{\circ (4)}$  \\  
$m_2/m_1$  &       a          &   b              &                &                        & (deg)            
}
\startdata 
50.0000    &   38.8473  & 24.1549 & -0.804949 &   0.403620  &  41.6369  \\ 
30.0000    &   30.4367  & 23.4695 & -0.762810 &   0.345029  &  46.2916  \\ 
27.0000    &   28.8461  & 23.3210 & -0.752768 &   0.333168  &  47.2562  \\ 
25.0000    &   27.6724  & 23.2126 & -0.745097 &   0.324583  &  47.9583  \\ 
22.0000    &   26.0610  & 23.0201 & -0.731694 &   0.310481  &  49.1218  \\ 
20.0000    &   24.6615  & 22.8834 & -0.721126 &   0.300119  &  49.9829  \\ 
17.0000    &   22.4351  & 22.6447 & -0.701887 &   0.282760  &  51.4453  \\ 
16.0000    &   21.7397  & 22.5510 & -0.694292 &   0.276412  &  51.9829  \\ 
15.0000    &   21.0134  & 22.4509 & -0.685943 &   0.269718  &  52.5555  \\ 
12.5000    &   18.7703  & 22.1764 & -0.660795 &   0.251257  &  54.1526  \\ 
10.0000    &   16.3794  & 21.8330 & -0.626603 &   0.229674  &  56.0519  \\ 
8.50000    &   14.7875  & 21.5831 & -0.599098 &   0.214695  &  57.4010  \\ 
8.00000    &   14.2367  & 21.4900 & -0.588236 &   0.209297  &  57.8883  \\ 
7.00000    &   12.7447  & 21.2984 & -0.563098 &   0.197727  &  58.9509  \\ 
6.66667    &   12.2733  & 21.2272 & -0.553484 &   0.193619  &  59.3329  \\ 
5.50000    &   10.7305  & 20.9442 & -0.513247 &   0.178097  &  60.7887  \\ 
4.50000    &   8.94808  & 20.6732 & -0.467151 &   0.163037  &  62.2355  \\ 
4.00000    &   7.88968  & 20.5249 & -0.438075 &   0.154758  &  63.0419  \\ 
3.33333    &   6.53351  & 20.3029 & -0.390097 &   0.142764  &  64.2243  \\ 
3.00000    &   5.81271  & 20.1833 & -0.360743 &   0.136275  &  64.8757  \\ 
2.00000    &   2.90722  & 19.8145 & -0.237418 &   0.114279  &  67.1305  \\ 
1.66667    &   1.66566  & 19.6972 & -0.177342 &   0.105850  &  68.0150  \\ 
1.33333    &   0.14992  & 19.6010 & -0.101000 &   0.096674  &  68.9854  \\ 
1.29870    &   0.03820  & 19.5908 & -0.091844 &   0.095658  & 69.1014  \\ 
1.29032    &  -0.02100  & 19.5895 & -0.089588 &   0.095428  & 69.1163  \\ 
1.25000    &  -0.16319  & 19.5787 & -0.078506 &   0.094236  &  69.2469  \\ 
1.11111    &  -1.03684  & 19.5608 & -0.037159 &   0.090024  &  69.6983  \\ 
1.00000    &  -1.54800  & 19.5482 &  0.000000 &   0.086509  &  70.0732  \\ 
0.750000   &  -3.28001  & 19.5878 &  0.101000 &   0.078010  &  70.9594  \\ 
0.599988   &  -4.84861  & 19.6866 &  0.177349 &   0.072363  &  71.5456  \\ 
0.500000   &  -6.17680  & 19.8050 &  0.237418 &   0.068302  &  71.9329  \\ 
0.333333   &  -9.34664  & 20.1759 &  0.360743 &   0.060609  &  72.6460  \\ 
0.300000   &  -10.1587  & 20.2907 &  0.390097 &   0.058871  &  72.7878  \\ 
0.250000   &  -11.7016  & 20.5085 &  0.438075 &   0.056050  &  73.0183  \\ 
0.222222   &  -12.8940  & 20.6642 &  0.467151 &   0.054348  &  73.1470  \\ 
0.181818   &  -14.9364  & 20.9407 &  0.513247 &   0.051642  &  73.3225  \\ 
0.150000   &  -16.7606  & 21.2100 &  0.553484 &   0.049224  &  73.4712  \\ 
0.142857   &  -17.3286  & 21.2829 &  0.563098 &   0.048631  &  73.5107  \\ 
0.125000   &  -19.0387  & 21.4884 &  0.588236 &   0.047066  &  73.5947  \\ 
0.117647   &  -19.7009  & 21.5779 &  0.599098 &   0.046382  &  73.6193  \\ 
0.100000   & -21.6154   & 21.8218 &  0.626603 &   0.044584  &  73.7114  \\ 
0.066667   & -27.2200   & 22.4429 &  0.685943 &   0.040421  &  73.8607  \\ 
0.062500   & -28.1263   & 22.5387 &  0.694292 &   0.039794  &  73.8775  \\ 
0.020000   & -49.9464   & 24.1552 &  0.804949 &   0.029898  &  74.0799   \\ 
\enddata
{\bf (1)} 
$\theta_s$ is the angle the $L1$ stream is making with the line joining
the two masses as the stream leaves $L1$. 
To a first approximation $\theta_s \approx (x-x_{L1})a+b$.
The value that has to be compared to \citet[][Table 1]{lub75} is $b$.  
{\bf (2)} $x_{L1}$ the x-coordinate of the $L1$ point.  
{\bf (3)} $\varpi_{\rm d}$ is the circularization radius,  
and {\bf (4)} $\varphi_{\rm h}-180^{\circ}$ is the circularization angle 
\citep[in Table 2 in][]{lub75}.  
\end{deluxetable}

\clearpage

\section{\bf Bow Shock Deflection} 

We now concentrate on the interaction of the incident $L1$ stream 
with the rim of the disk. 
While we consider only the region above the disk, namely
for $z \ge 0$, a symmetrical
picture holds below the disk for $z \le 0$. 
As the stream leaves the $L1$ region, its interaction with the 
rim of the accretion disk can be summarized as follows: 

\indent
{\small{{\bf(0)}}} 
Because both the stream and disk are highly hypersonic, a bow shock 
develops at the impact region \citep{lub76,arm98,kun01}. 

\indent 
{\bf (i)} 
At an elevation $z$ high enough above the disk plane 
(i.e. for $z > H_0$, where $H_0$ is larger than the disk vertical
scale height $H$),  the $L1$ stream density 
becomes significantly larger than that of the disk, and stream material  
is able to pass above the disk without being  affected
by its presence 
\citep{lub75,lub89}. 

\indent 
{\bf (ii)} 
Within the context of a hypersonic flow past a blunt object, we expect  
the stream matter to be deflected upward across the  
bow shock \citep{rat10}. This is somewhat similar to the simulations of the
isothermal case in \citet{arm98} and the $\gamma \approx 1$ polytropic case in
\citet{kun01}.   

\indent 
{\bf (iii)} 
At higher mass accretion rates, and/or if the $L1$ stream velocity 
is large, a heated region will develop in the shock, first in the 
vicinity of the disk midplane where the density is larger,  then 
possibly in the entire shock region if cooling becomes inefficient. 
Here we consider the case in which the midplane region of the shock
might be heated, but higher up, the shock is still behaving 
in a manner similar to the isothermal case.    

\indent 
{\bf (iv)} 
Eventually, as cooling becomes inefficient, the entire shock region
becomes hot enough to expand and stream matter might even be ejected
due to adiabatic expansion. This is similar to the adiabatic case
in \citet{arm98} and the larger polytropic coefficient simulations of 
\citet{kun01} in which more stream matter is stopped at the hot spot
region. One can expect here the hot spot to extend vertically and 
laterally and to even form a ``tail''.  
In this case it is likely that a hot bulge forms (at the hot spot location) 
splashing in all directions without the  formation of a coherent stream
\citep{kun01}. 

In the present work, we concentrate only on the cases that can be approximated 
within the context of a nearly isothermal hypersonic flow past a blunt 
object forming a bow shock, namely cases (0), (i), (ii), and (iii).   
We do not consider additional scenarios 
such as for example a tilted disk, where the stream 
might completely miss the rim; or a thick disk impacted by a thin stream
without overflow.

\subsection{{\bf The Stream-Disk Interaction as a Hypersonic Flow past a Blunt Object}} 

Since our main assumption rests on the hydrodynamical problem of the  
hypersonic flow past a blunt object, we consider now the 
general behavior of such a flow, with the $L1$ stream as 
the hypersonic flow and the disk as the blunt object.  
The general problem has been extensively studied analytically, numerically 
and in wind tunnel experiments, since it directly applies to the flight 
of a rocket/missile in the Earth's atmosphere \citep[e.g.][]{rat10}.  

\clearpage

\begin{figure} 
\vspace{-8.cm} 
\plotone{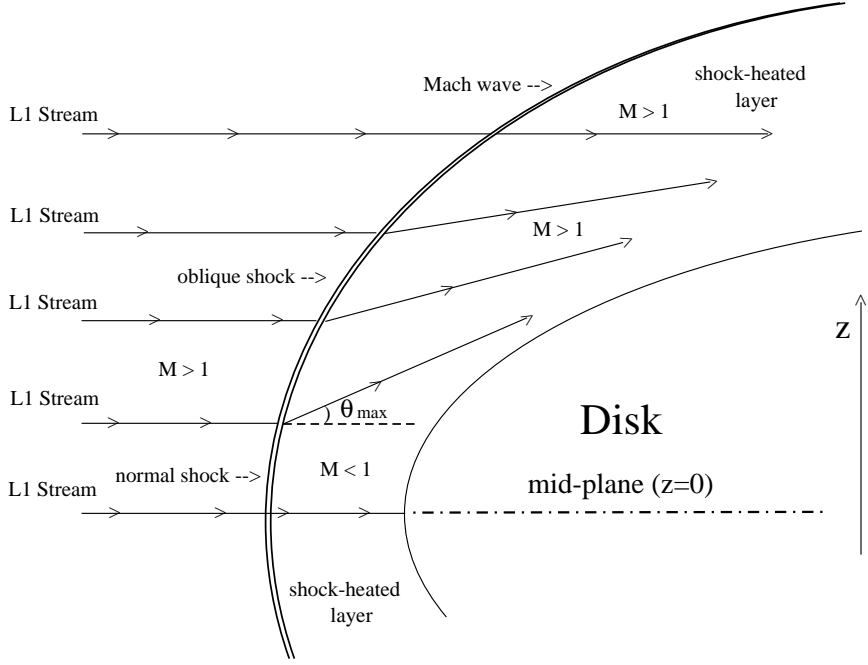} 
\vspace{-1.cm} 
\caption{Schematic illustration of the $L1$ stream impacting the 
outer edge of the disk forming a bow shock.
The hypersonic ($M>>1$) 
$L1$ stream flows in from the left and impacts the disk located
on the right. For clarity,  only the $z>0$ region is shown. 
At the stream-disk interface a bow shock 
forms, shown here with the double black line. By analogy with the hypersonic
flow past a blunt body, one can differentiate between three 
regions. 
{\bf(1)} 
Near the mid-plane of the disk (with a higher density) 
the shock is normal to the velocity and the 
downstream velocity is subsonic. 
{\bf (2)} Higher up, the shock is oblique and the 
downstream velocity changes gradually from subsonic (strong shock) 
to supersonic (weaker shock) as $z$ increases.  
{\bf (3)} At still higher elevation, as the shock weakens, 
it forms a Mach wave and the downstream velocity is hypersonic.
The $L1$ stream material in region (3) 
continues mostly unaffected by the shock; 
at lower elevations, through the 
oblique shock in region (2) and the normal shock in region (1), 
the material is deflected upwards to a maximum angle
$\theta_{\rm max}$. 
}
\end{figure} 

We represent such a configuration with the disk
seen edge-on in Fig.5.  
The $L1$ stream flows in from the left and impacts the disk located
on the right. 
As a result a ``detached'' shock forms in front of the disk edge,  
in the shape of a bow: the {\it bow shock}. 
In the region near the disk midplane ($z \approx 0$), 
the shock is perpendicular (normal)  
to the velocity of the stream and the shock is the strongest there.  
Downstream from the normal shock, in the stagnation region 
(i.e. between the normal shock and the disk), 
the velocity is subsonic. 
As one moves away from the midplane along the bow shock 
(as $z$ increases),  
the shock becomes oblique  and the downstream velocity changes gradually 
from subsonic (strong shock) to supersonic (weaker shock). 
Downstream from that region the flow is deflected upward.  
Farther away from the midplane, as the shock weakens, 
it forms a {\it Mach wave} (a weak shock)
and the downstream velocity remains hypersonic.
The $L1$ stream material high above the disk midplane, in the Mach wave
region, continues mostly unaffected by the shock; 
at lower elevations, through the 
oblique shock and the normal shock, 
the material is deflected upward to a maximum angle
$\theta_{\rm max}$. 
Across the bow shock, there is no change in the velocity component 
tangent to the shock; however, the velocity component
normal to the shock is decelerated to a subsonic value.

\clearpage

\subsection{{\bf Bow Shock Deflection off the Disk Plane}}  

The vertical deflection of the 
the $L1$ stream by the bow shock can be formulated 
as follows.  
We simulate the trajectory of test particles from $L1$, 
assuming that the stream has a vertical extent ($z>0$) but  
flows horizontally ($V_z =0$) until it reaches the bow shock. 
At a given height, a test particle impacts the bow shock at a point
${\rm P}$, where the bow shock surface makes an angle $\theta$
with the plane of the disk (XY plane; see Fig.6, left panel).  
There, the component of the velocity normal to the shock surface is
reduced to a subsonic value, while the component parallel to the
shock remains unchanged. 
The particle velocity before impacting the disk $V=V_x^2+V_y^2$ 
is decomposed (Fig.6, right panel) into a
component parallel to the direction of the deflection (making an
angle $\theta$ with the $XY$-plane) $V_{\|}$ 
and a component perpendicular
to the shock (making an angle $-(\pi/2-\theta)$ with
the $XY$-plane) $V_{\bot}$.

We assume that before impact the velocity is hypersonic, such
that $V = {\cal{M}} c_s^0$, where $c_s^0$ is the sound speed before
impact (upstream), and ${\cal{M}} >>1$. For example, \citet{arm98} assumed  
${\cal{M}}=30$ in their isothermal case simulation, and we adopt
here this value.   
After the shock, the perpendicular component $V_{\bot}$ of the  
velocity is set to subsonic; $V_{\bot} = \beta c_s^1$ with $0 < \beta < 1$
and $c_s^1$ is the sound speed after the shock (downstream).
In the present case we chose $\beta =0.5$.
Therefore, before impact one has: 
\begin{equation} 
\left\{
\begin{array}{l} 
V_{\|}=V \cos{\theta},   \\ 
V_{\bot}=V \sin{\theta} , \\ 
V = {\cal{M}} c_s^0 .  
\end{array}
\right. 
~~~~{\rm and}~~~~~~
\left\{
\begin{array}{l} 
V_x= V \cos{\varphi_0} ,  \\ 
V_y= V \sin{\varphi_0} , \\   
V_z= 0 , 
\end{array}
\right.
\end{equation}   
where $\varphi_0$ is the angle the velocity $\vec{V}$ is making 
with the X-axis (see Fig.7).  
After impact the velocity components become\footnote{strictly speaking
$ V_{\bot}=   {\cal{M}} c_s^0 \sin{\theta}$ when 
$ {\cal{M}} c_s^0 \sin{\theta} < \beta c_s^1$, which happens for 
$\theta \approx 0$. However, within the approximations we are making
here (see also the discussion), this is negligible.   
}:  
\begin{equation} 
\left\{
\begin{array}{l} 
V_{\|}= {\cal{M}} c_s^0  \cos{\theta},   \\ 
V_{\bot}=   \beta c_s^1  , \\
0 < \beta < 1 .\\   
\end{array}
\right.
\end{equation}   


\begin{figure}[b!]
\vspace{-7.cm} 
\plottwo{f06a.eps}{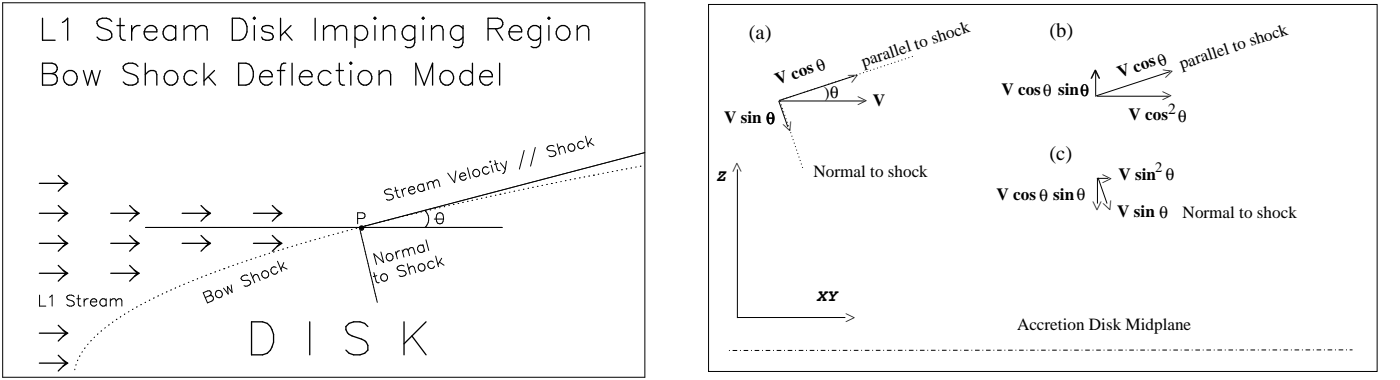}  
\vspace{-3.cm} 
\caption{Schematic model of the stream 
from the first Lagrangian point, $L1$, deflected
{\it across} the bow shock. {\bf Left.} 
The $L1$ stream material is moving from the left with a velocity $V$
and collides with the disk forming a bow shock.  
As a result, at any point $P$  
matter is deflected {\it across} the bow shock 
and launched on a ballistic trajectory. 
The surface of the bow shock at $P$ forms an angle $\theta$ 
with the plane of the disk (horizontal). 
At the front of the shock, lower left, $\theta=90^{\circ}$,
downstream to the right one has $\theta \rightarrow 0$.  
{\bf Right.} The incoming (test) particle velocity components 
are decomposed.
}
\end{figure}

\clearpage

We decompose $V_{\|}$ back into its $x,y,z$ components (Fig.6, right panel). 
Across the hot spot region, we therefore have 
\begin{equation} 
\left\{
\begin{array}{l} 
V_x \rightarrow (c_s^0  {\cal{M}} \cos^2{\theta} 
+ c_s^1 \beta \sin{\theta} ) \cos{\varphi_0}  , \\   
V_y \rightarrow (c_s^0  {\cal{M}} \cos^2{\theta}    
+ c_s^1 \beta \sin{\theta} ) \sin{\varphi_0}  , \\   
V_z \rightarrow (c_s^0  {\cal{M}} \sin{\theta} - c_s^1 \beta ) \cos{\theta} . \\   
\end{array}
\right.
\end{equation} 
If we now further assume that the shock is nearly isothermal, we have 
$c_s^1 \approx c_s^0$ and 
\begin{equation} 
\frac{c_s^1 \beta}{c_s^0 \cal{M}} = \frac{0.5}{30} << 1 .
\end{equation} 
We therefore drop the $c_s^1 \beta$ term to simplify the expression
(see also the discussion section).
The velocities  are reduced to:
\begin{equation} 
\left\{
\begin{array}{l} 
V_{\|}=V \cos{\theta},   \\ 
V_{\bot}=V \sin{\theta} \rightarrow 0, 
\end{array}
\right.
\end{equation}   
and decomposing $V_{\|}$ back into its $x,y,z$ components, 
we obtain    
\begin{equation} 
\left\{
\begin{array}{l} 
V_{x} \rightarrow V_{x} \cos^2{\theta} , \\   
V_{y} \rightarrow V_{y} \cos^2{\theta} , \\   
V_{z}:~ 0 \rightarrow \sqrt{V_{x}^2+V_{y}^2} \cos{\theta} \sin{\theta}. \\   
\end{array}
\right.
\end{equation} 
Under this assumption, 
the velocity perpendicular to the shock is simply set to zero,
and the angle the shock is making with the XY plane, $\theta$, becomes
the angle at which the particles are launched off the disk plane. 
Within the XY (disk) plane, the velocity of the particle keeps the same
direction. For $\theta=0$, the particle velocity remains unchanged as it
passes above the disk; for $\theta=\pi / 2$, the particle velocity 
goes to zero as it impacts the disk mid-plane.   
Preliminiary results for this assumption {\it only} were presented
for EM Cyg in \citet{god09}.

\subsection{{\bf Second Assumption: Matching the Disk Keplerian Velocity} }

As the stream matter impacts the disk edge, in addition to being 
deflected off the disk plane, it is expected
that the radial (directed from the WD) component of the stream velocity will be reduced
or even possibly canceled, at least in these regions 
where the disk density is much larger than the stream density. 
At the disk midplane the radial component of the stream velocity must
vanish, while at a few vertical disk scale heights $H$ the stream velocity 
(and direction) will not be affected at all. 

Let us decompose the stream velocity $(V_x,V_y)$, before impact, in its
tangential $V_T$ and radial $V_R$ components (see Fig.7), 
\begin{equation}
\left\{
\begin{array}{l}
V_R = -V_x \cos{\phi} - V_y \sin{\phi} , \\  
V_T=V_x \sin{\phi} - V_y \cos{\phi} ,  
\end{array}
\right.
\end{equation}
where the angle $\phi$ defines the place where the stream hits the
edge of the disk as viewed from the primary, in the disk plane, and it 
is measured counterclockwise from the line joining the two stars.

For stream matter impacting the disk edge at the midplane
the radial component of the velocity vanishes 
completely, $V_R \rightarrow 0$,
while for stream matter at several scale height $H$ above
the disk $V_R$ remains unchanged, as the stream flow is not
interacting with the disk. 
Similarly, we assume that the tangential velocity $V_T$ will
match the Keplerian velocity $V_K$ at the disk edge for $\theta=\pi/2$ 
and will remain unchanged for $\theta=0$. 
We therefore introduce a
function  $\alpha  $ of $\theta$, such that 
\begin{equation}
\left\{
\begin{array}{lcl} 
\alpha (\theta) \rightarrow 0 & for  & \theta \rightarrow \pi/2 , \\
\alpha (\theta) \rightarrow 1 & for  & \theta \rightarrow 0 , \\
0 < \alpha(\theta) < 1 & for & 0 < \theta < \pi/2. 
\end{array}
\right. 
\end{equation} 
With this definition, the velocity components are transforming as
follows:  
\begin{equation}
\left\{
\begin{array}{l}
V_R \rightarrow \alpha (\theta) V_R  , \\ 
V_T \rightarrow \alpha (\theta) V_T +[ 1 - \alpha ] V_K ,  
\end{array}
\right.
\end{equation}

\begin{figure}[b!]
\vspace{-5.cm} 
\gridline{ 
           \fig{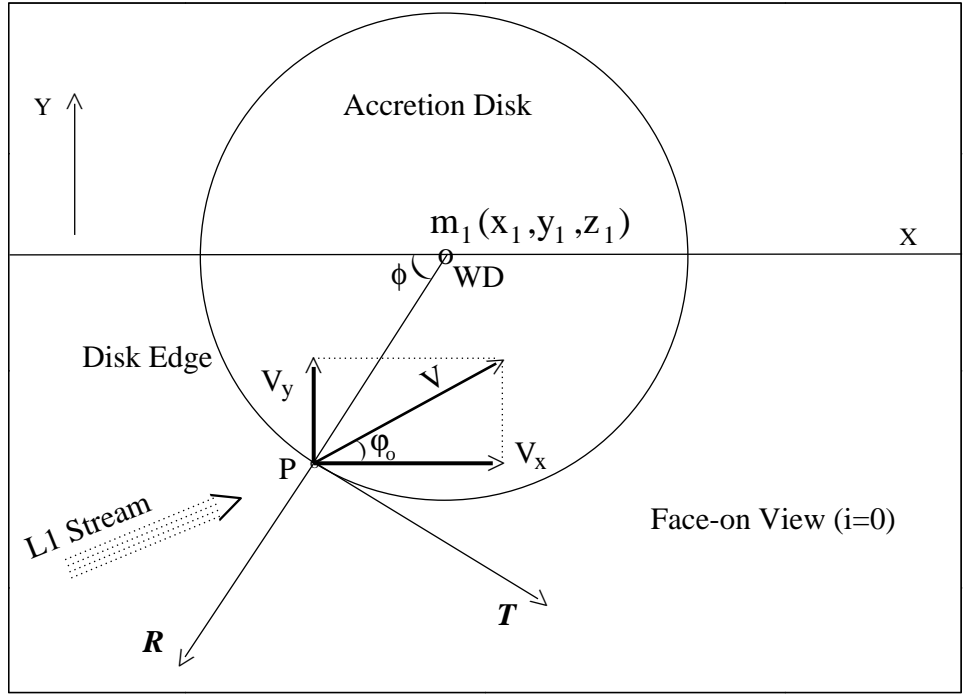}{0.8\textwidth}{}
          }
\vspace{-2.cm} 
\caption{
Model of the $L1$ stream deflection in the XY plane. 
The $L1$ stream velocity, at the disk edge impact,
makes an angle $\varphi_0$ with the X-axis. 
We adopt the notation $\phi $ for the angle at which 
the stream impacts the disk edge. 
}
\end{figure}

Choosing $\alpha(\theta) = \cos^2{\theta}$, from the previous subsection
(equation 12, see also the Discussion section), gives:  
\begin{equation}
\left\{
\begin{array}{l}
V_R' = V_R \cos^2{\theta}  , \\ 
V_T'=  V_T \cos^2{\theta} + V_K \sin^2{\theta} ,  
\end{array}
\right.
\end{equation}
where the prime denotes the velocities after deflection. 
%
%
After the stream impacts the disk, the $(x,y,z)$  
components of the velocity are given by the final expressions
\begin{equation} 
\left\{ 
\begin{array}{l} 
V_x' = V_x \cos^2{\theta}  + V_K \sin^2{\theta}\sin{\phi} , \\  
V_y' = V_y \cos^2{\theta} - V_K \sin^2{\theta} \cos{\phi} , \\   
V_z = V \cos{\theta} \sin{\theta} . 
\end{array} 
\right.
\end{equation} 

For $\theta \rightarrow 0$ the
particle trajectory is not deflected as it does not interact with
the edge of the disk; rather it passes above the disk. For $\theta \rightarrow \pi/2$, 
the particles impact the mid-plane of the disk; 
it undergoes the largest change in its velocity, losing
its inner radial (toward the WD) component and a gaining a Keplerian
component in the tangential direction (e.g. motion in the disk around the WD;
assuming that the outer radius of the disk is not too large; otherwise, the secondary
gravitational potential has to be taken into account). 

\clearpage

\section{{\bf Simulations \& Results}}

For each $L1$ stream particle (with index $i=0,1,..,90$) we simulate the impact 
with the disk by launching it on a ballistic trajectory from a   
point $P_i (R_i, z_i)$, where $z_i =H_0 - \Delta z_i$, $\Delta z_i = H_0 (i/90)$, 
and $R_i = R_{\rm disk} - \Delta z_i$, with a launching angle 
$\theta = i^{\circ}$ (see Fig.8). 
$H_0$ is the minimum height ($z$) at which the $L1$ stream flows unaffected 
by the disk - a measure of (half) the vertical thickness of the bow shock.
$H_0$ is unknown; it is likely of the order of a few disk scale heights $H$.
In the simulations, it is varied until the data fit the observations.  
Our choice of  $\Delta z_i $ 
is consistent with the assumption that the $L1$ stream flow is not deflected
if $z > H_0$.  
While the launching angle takes values from $0^{\circ}$ to $90^{\circ}$, 
the points $P_i$ are lined up on a curved diagonal, rather than an 
hyperbolic. We found that the exact shape (curved diagonal vs hyperbolic) 
of the curve has no noticeable effects on the results. 
Since $ H_0/R \sim$ a few $H/R$, itself just a few percent, the effect of 
the curve defining $P_i$ is very small as long as the post-shock velocity
is dictated by the angle of deflection. 
The velocity imparted to each particle is then given by equation (17) above
with $\theta = 0^{\circ}$ to $90^{\circ}$.  
The particle is then followed until it falls back onto the disk at $z \approx H$. 
In the limit $\theta \rightarrow 0$ and $\Delta z \ne 0$ 
we obtain the same result as \citet{lub89}, namely that the particle falls
back onto the disk around orbital phase 0.5-0.6 (see Fig.9). 
For $\theta \ne 0$ we obtain 
that the particle is launched and falls
back onto the disk at an angle $\phi_{impact}> 150^{\circ}$. As 
$\theta$ increases, the particle reaches a maximum height above the disk,
while the impact point on the disk reaches 
a maximum angle of $\approx 180^{\circ}$ (see Figs.8 \& 9).

\clearpage

\begin{figure}
\vspace{-4.0cm} 
\gridline{\fig{f08a.eps}{0.5\textwidth}{}
          \fig{f08b.eps}{0.5\textwidth}{}
          }
\vspace{0.7cm} 
\caption{
Simulation of the bow shock deflection launching of the $L1$ stream material
at the outer edge of the disk. 
On the {\bf left} we show a nearly edge-on view, and on the  
{\bf right} we show a top view.  
7 ballistic trajectories, for test particles moving from left to right, 
are shown in black,  the accretion disk edge is in red.
The axis are in units of the binary separation centered on the primary WD star 
- the red circle (not to scale) located at $(0,0)$. 
Trajectory {\it a} represents a particle passing at a height {\bf $z = H_0$}  
above the disk, which is not deflected at all, corresponding to a launching 
angle $\theta = 0$.  
Trajectory {\it g} represents a particle hitting the disk 
at $z=0$ and being dragged with the Keplerian flow, corresponding to 
a launching angle $\theta = 90^{\circ}$. 
For clarity we show only 7 partial trajectories, with launching angles
as follows: 
$\theta=0^{\circ}$ (a),  $\theta=15^{\circ}$ (b),  $\theta=30^{\circ}$ (c),  
$\theta=45^{\circ}$ (d),  $\theta=60^{\circ}$ (e),  $\theta=75^{\circ}$ (f),  
and $\theta=90^{\circ}$ (g). 
The trajectories do not intersect. 
In the configuration shown the mass ratio is $m_2/m_1=0.35$, the 
disk outer radius $R_{\rm disk}=0.376a$ ($a$ is the binary separation),  
and {\bf $H_0/R=0.10$}. 
} 
\vspace{-3.cm} 
\gridline{
          \fig{f09.eps}{0.5\textwidth}{} 
           }
\vspace{0.7cm} 
\caption{
Three-dimensional perspective view of the binary  at an inclination angle
of $70^{\circ}$ and orbital phase $\Phi=0$
(the secondary is in front).
The configuration is the same as in Figs.8, except 
that 19 particles are shown with deflection angles $\theta =0^{\circ}$, $5^{\circ}$, 
$10^{\circ}$,$15^{\circ}$, ..., $85^{\circ}$, and $90^{\circ}$. 
}
\end{figure}

\clearpage 

\subsection{{\bf U Gem Input System Parameters}}  

We now model the velocity offsets as a function of the orbital phase
for the centers of the absorption lines in the {\it FUSE} spectrum of 
U Gem taken during an outburst \citep{fro01}.
For that purpose, we review and discuss U Gem system parameters, as
they are used as input parameters in the simulations. 

\paragraph{Orbital Period and Masses.} 
U Gem has a period of 0.1769061911 days  
(4.24574858 hr; \citep{mar90,ech07,dai09}), 
and a mass ratio $q=0.35-0.362$ \citep{lon99,sma01,ech07}.
In the following, we scale the velocity offsets to the WD maximum
radial velocity (amplitude) both in the observations and
in our simulations. As a consequence, the WD
mass {\it per se} does not explicitly enter into our
simulation; it enters implicitly in the mass ratio.   
This has the advantage of being independent of the WD
mass, as its value could be anywhere between $1.05 M_{\odot}$
\citep[as inferred from scaling the UV flux to the known distance]{nay05,ech07} 
and $1.25 M_{\odot}$ 
\citep[derived from the gravitational redshift]{sio98,lon99,nay05,ech07}.
In the simulations we set $q=0.35$.

\paragraph{Inclination and Eclipsing Geometry.}
U Gem is a partially eclipsing binary \citep{krz65}, 
as the disk undergoes partial eclipses
(at phases 0.95-0.05; \citet*{ech07}), the hot spot undergoes full eclipses, 
while the WD itself is not eclipsed at all \citep{sma71,war71}.  
This restricts the inclination to be 
in the range $67^{\circ}$$^{+7^{\circ}}_{-5^{\circ}}$ \citep{lon99,lon06}, 
though \citet{lon99} adopted $67^{\circ} \pm 3^{\circ}$. 
A value $i=69.7^{\circ} \pm 0.7^{\circ}$ was first derived by \citep{zha87},
which has been reassessed since then. A low inclination ($\sim 65^{\circ}$)
is consistent with a large WD mass ($\sim 1.25M_{\odot}$), while a 
larger inclination ($\sim 73^{\circ}$, \citet{und06}) is consistent with a smaller 
WD mass ($\sim 1.05 M_{\odot}$). The larger WD mass has been corroborated 
by its gravitational redshift measurement, but the corresponding smaller
WD radius is inconsistent with the UV flux scaling to the distance of the system.  
For that reason, we consider inclinations from $i= 62^{\circ}$ to $i=75^{\circ}$. 

\paragraph{The Outer Radius of the Disk.} 
The outer radius of the disk is expected to be larger during
outburst than during quiescence \citep{sma01} such that  
eclipses of the hot spot are not always observed
\citep{mar90}, and the secondary undergoes partial eclipses 
by the disk of only a few percent \citep{sma71}.  
The hot spot eclipses undergo a phase shift and occur earlier  as the brightness rises;
when the systems fades, the eclipses return to the original phase \citep{krz65}.  
Since these are the eclipses of the hot spot by the secondary, 
this implies that the hot spot moves on the rim of the disk 
toward the secondary (retrograde motion) during the rise to
outburst and vice versa during decline. For this to happen, the disk 
outer radius has to grow during rise and it has to decrease during decline,  
and one then expects the radius of the disk during quiescence to 
be rather small.  However, 
\citet{ech07} report a symmetric {\it full} disk with 
an outer disk radius in deep quiescence of $0.61$ the binary
separation, or about twice as large as expected
(e.g. \citet{lub89}). Such a large disk radius 
almost reaches  the first Lagrangian point 
$L_1$ and is larger than the {\it average} 
Roche lobe radius itself.
The largest quiescent disk radius obtained by \citet{sma01}
was $\sim 0.5$ the binary separation, namely, 20\% smaller. 
It is likely that not all outburst cycles show the same behavior,
and consequently the outer radius of the accretion disk has to 
be considered as an {\it unknown} parameter. In the simulations, we vary the outer
radius of the disk from $\sim$0.20~a to $\sim$0.60~a.  

\paragraph{The Bow Shock Height $H_0$.}
The minimum height $H_0$, at which the stream flows unaffected by the disk, 
corresponds in our model to a measure of the vertical thickness of the bow shock. 
$H_0$ is possibly of the order of a few disk scale heights $H$ and is 
taken as an additional unknown parameter. We vary $H_0/r$ from 
0.02 to 0.30. 

\subsection{{\bf Output: Particle Coordinates and Velocities}}

Our restricted three-body code solves the particles coordinates and velocities as a 
function of time. The coordinates as a function of time provide 
the trajectories of the particles in three dimensions. 
We carry out a transformation of the trajectories (paths) of the particles
into an inertial frame of reference
as the three-dimensional projection 
viewed by the observer for a given orbital inclination $i$ and 
orbital phase $\Phi$. An example is given in Fig.10, where, for 
convenience, the graph and its axis in each panel are centered on the WD.

\clearpage

\clearpage 

\begin{figure}
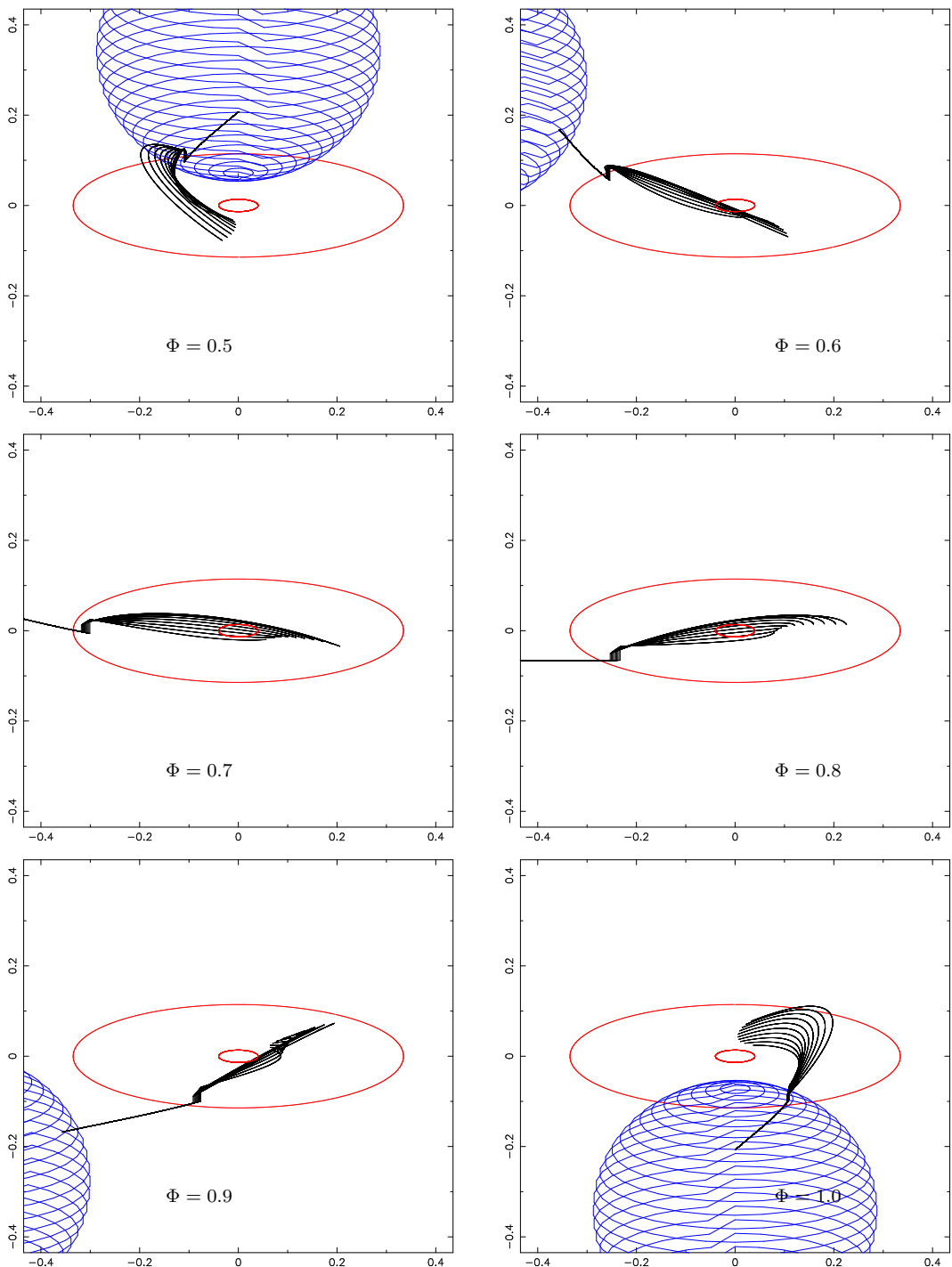

\vspace{-5.0cm} 
\gridline{\fig{f10a.eps}{0.50\textwidth}{$\Phi=0.5$}
          \fig{f10b.eps}{0.50\textwidth}{$\Phi=0.6$}
} 
\vspace{-4.5cm} 
\gridline{\fig{f10c.eps}{0.50\textwidth}{$\Phi=0.7$}
          \fig{f10d.eps}{0.50\textwidth}{$\Phi=0.8$}
} 
\vspace{-4.5cm} 
\gridline{\fig{f10e.eps}{0.50\textwidth}{$\Phi=0.9$}
          \fig{f10f.eps}{0.50\textwidth}{$\Phi=1.0$}
} 
\vspace{1.0cm} 
\caption{
Perspective view of the 
binary configuration with the $L1$ stream launched above the disk
at phase $\Phi=0.5$ (upper left), 0.6 (upper right and so on), 
0.7, 0.8, 0.9, and 1.0. The secondary is in blue, the stream is in 
black and the accretion disk outer edge is represented with the
large red circle. The smaller inner red circle delimits the  
inner region of the disk that is being veiled by the stream: 
the radial velocities of the stream
test particles are only being considered inside this region
(see text).
Only 10 ballistic trajectories are being considered in this 
figure, with launching angles 
$\theta = 0^{\circ}$, $5^{\circ}$, $10^{\circ}$,..,$40^{\circ}$, $45^{\circ}$.
In the simulation shown in the figure,  
the veiling of the inner disk by the stream material begins
a little before phase $\Phi = 0.6$ (the {\it ingress}), 
and ends around phase $\Phi \approx 0.9$ (the {\it egress}).  
}
\end{figure}

\clearpage

The {\it FUSE} spectrum of U Gem, taken while the disk dominated the FUV,
has a wavelength range of $\sim 910-1185$~\AA . 
This implies that 
the inner accretion disk (the hottest region of the disk) 
is the predominant region contributing to the {\it FUSE} spectrum,   
while the outer disk emits in the optical, and the intermediate region
emits mainly in the near-UV.  
As a consequence, it is the {\it veiling of the inner disk} by the $L1$ stream
material that is responsible for the velocity offsets of the metal
absorption lines as measured in the {\it FUSE} spectrum. 
We delimit the inner disk region with a radius $R_v$ (the {\it veiling radius}),
and we only consider particles veiling the disk in the region 
$R < R_v$ (for a given inclination and as a function of the orbital phase,
see Fig.10). 
The size of the veiling radius $R_v$ is a {\it variable to be found}.  
 
Each point on a trajectory has a corresponding velocity, and we therefore
consider the velocity of these partial trajectories within $R <R_v$.
For each individual particle (each with a different launching angle $\theta$), 
we average the projected (to the observer) radial velocity over the partial
trajectory delimited within $R<R_v$.    Doing so, we obtain for each
particle (i.e. for each angle $\theta$) 
an average radial velocity as a function of the orbital phase
$\phi$ (for a given inclination). 
 
In addition, we have to consider the fact that at some specific phases,
such as e.g. 0.6 and 0.9 in Fig.10, the stream veils only a fraction
of the inner disk. These are transition regions, similar to eclipse 
ingress and egress,  and we do not expect the stream material to generate 
noticeable absorption lines if it covers less than 50\% of the inner
disk.

\subsection{{\bf Results}}  

We ran many simulations varying the parameters as described in the
previous section, and we present the results  
by drawing the particle average radial velocity 
(see previous subsection)
as a function of the orbital
phase for direct comparison with the velocity offsets of metal lines of U Gem. 
For that purpose, Froning et al's data have been slightly improved. 
Namely, the WD velocity has been adjusted to match  
the slightly larger amplitude as derived in \citet{lon06}. 
We also used the latest value of the ephemeris \citep{ech07} 
and included heliocentric
corrections that increase the orbital phases of the all the observed data points 
by $\sim 0.05$.  
The velocities have all been normalized to the maximum
projected orbital velocity of the WD ($=1$). 

For each simulation, the model is drawn on top of the observation
as shown in Fig.11.   
We carry this procedure while changing the orbital inclination $i$, the outer
radius of the disk $R_d$, and the thickness {\bf $H_0/r$}.
For each such simulation ($i$, $R_d$, {\bf $H_0/r$}), we check the results 
for a range of launching angles $\theta$ (i.e. for a number of particles) 
and veiling radius $R_v$. In this procedure we try to match the
velocity offset peak ($V_{\rm off}=5.2)$,  the orbital phase of the peak 
$\Phi \approx 0.72-0.77$, and the velocity offset drop observed
at phase $\Phi=0.81-0.82$.    
In Fig.11 we show two models with $i=66^{\circ}$ and $i=70^{\circ}$
to illustrate our technique. The model with the higher inclination
produces a velocity maximum in excess of 6, and in both models
 the velocity peaks a little after phase $\Phi=0.8$.  
The phase at which the velocity offset peaks
depends strongly on the outer radius of the disk, and the maximum
velocity decreases with decreasing inclination and decreasing 
thickness {\bf $H_0$}. 

Consequently, 
we are able to better match the data by decreasing the disk radius 
while keeping
the orbital inclination at $i=66^{\circ}$. We present this model in
Fig.12. In this model {\bf $H_0/r=0.08$}, and the maximum launching angle
we obtained is $\theta=43^{\circ}$. 
The need to keep $\theta \le 43^{\circ}$ is due to the minimum
velocity ($\sim 3.69$) observed at phase $\Phi = 0.72-0.77$:
for $\theta > 43^{\circ}$ the model produces lower velocity at
that phase. The veiling radius is $R_v=0.025a$, it also dictates the
lowest velocity, and especially around phase 0.8 at which 
the velocity drop is observed. A significantly larger veiling radius 
is ruled out because it would 
include particles with lower velocities, adding theoretical 
data points to the model
with radial velocity of $\sim 3$ at phase $0.7-0.8$
(as seen in Fig.11 for $i=66^{\circ}$), which is not seen in the
observational data.

\clearpage

\begin{figure}
\vspace{-5.cm}
\plotone{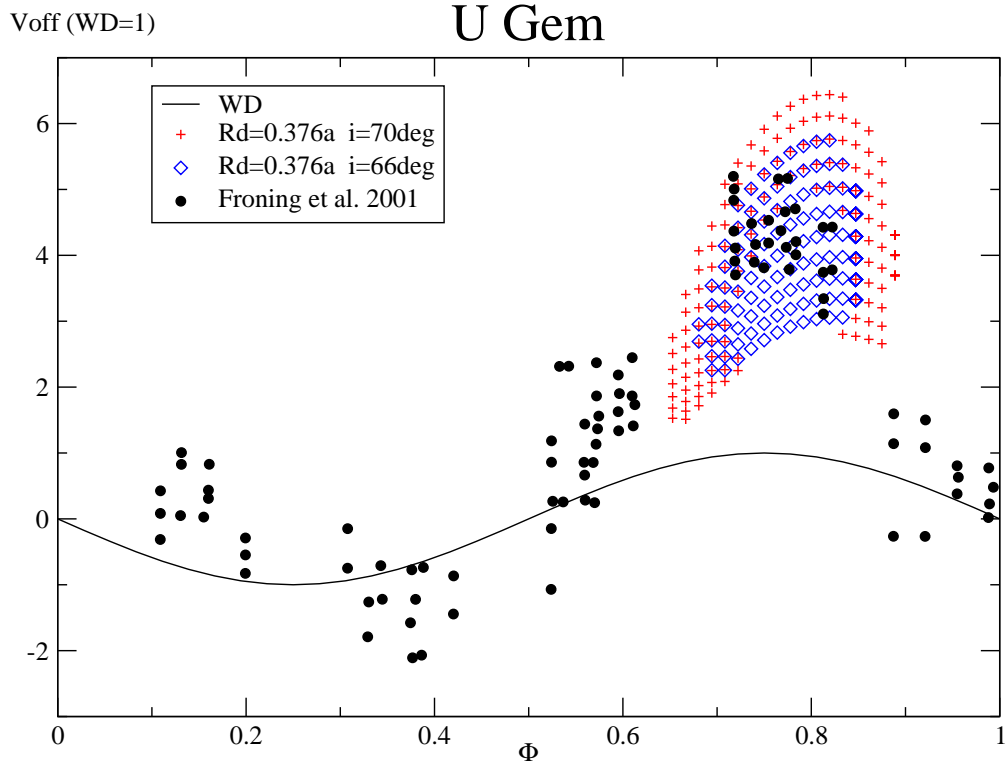}
\caption{
Velocity offsets of metal absorption lines in the {\it FUSE} spectra
of U Gem obtained in outburst as a function of the orbital phase
(from \citet{fro01}, as in Fig.3d, all the circles are now filled),  
together with two ballistic trajectory models. 
Two stream-disk overflow ballistic trajectory 
simulations are shown for comparison assuming 
a relatively large outer radius for the disk, $R_d=0.376a$, 
{\bf $H_0/r=0.10$}, and a binary inclination of $i=70^{\circ}$ (red crosses) and 
$i=66^{\circ}$ (blue diamonds). 
The lower inclination agrees better with the a maximum velocity
of $\sim 5-6$, but both models are off-phase by $\sim 0.05-0.10$.   
Note that between phase $\sim 0.6-0.7$ there are no data points
for comparison. 
} 
\end{figure}

\clearpage 

\begin{figure}
\vspace{-5.cm}
\plotone{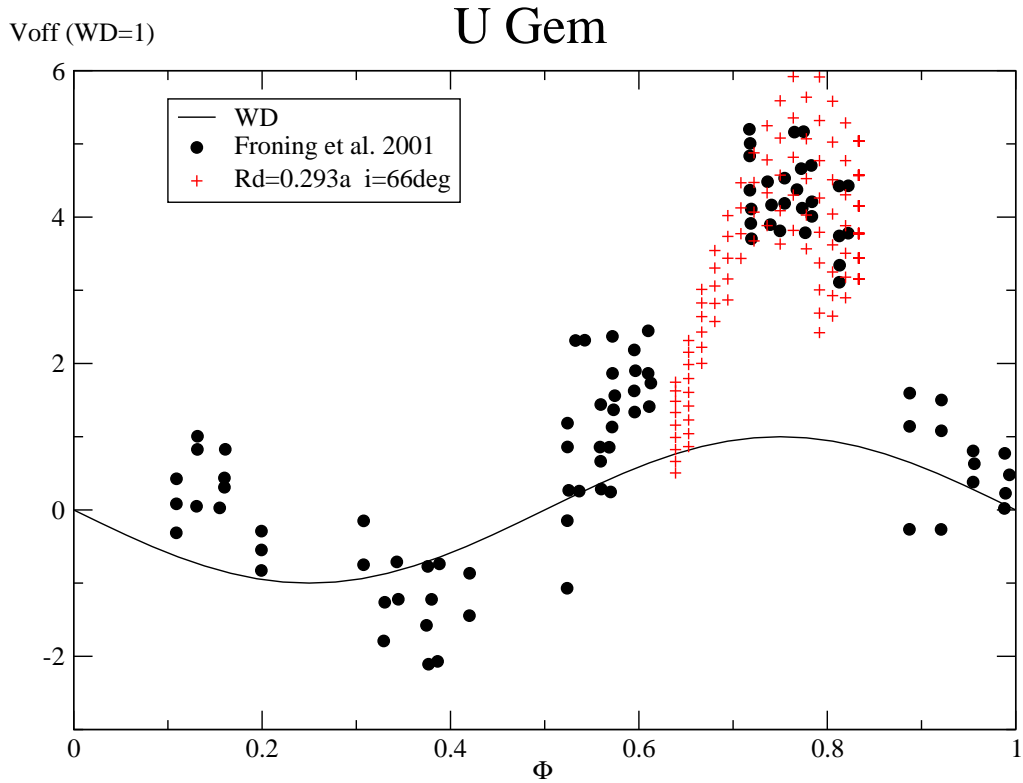}
\caption{
Velocity offsets of metal absorption lines (filled black circles)  
of U Gem in outburst as a function of the orbital phase as in the previous
figure, together with a ballistic trajectory model (red crosses). 
Matching the observed data 
is obtained by decreasing the inclination of the model to 
$i=66\pm 1^{\circ}$ and setting the outer disk radius to 
$R_d=0.29 \pm 0.023a$. The model is able to reproduce the drop in 
velocity seen around phase $\sim 0.8$. In this model {\bf $H_0/r=0.08 \pm 0.02$} and the
portion of the inner disk that is masked by the stream is within
$r \le 0.025 \pm 0.005a$. 
At the edge of the disk the test particles 
are launched at a deflection angle of up to $43^{\circ}$. 
} 
\end{figure}

\clearpage 

\section{{\bf Discussion}} 

Overall, our results are able to reproduce the velocity shift of the 
absorption lines and are consistent with some of the system parameters
(i.e. inclination, mass ratio,etc..). However, some of the approximations
and assumptions we made are crude (e.g., isothermal) while other parameters  
(e.g. thickness $H_0$, outer radius) are actually unknown. 
We therefore need to address the values of the parameters we chose 
and how they affect the results, and check whether our assumptions
are self-consistent. 

\paragraph{Bow Shock Thickness $H_0$.}  
We obtained a bow shock (half) vertical thickness of $H_0/r=0.08 \pm 0.02$, 
which was found by fitting the maximum velocity offset  
and its phase. Since $H_0$ is of the order of a few $H$,  
this result is consistent with the thin-disk approximation 
\citep{pri81} in the standard disk theory. 
If we consider a Keplerian disk with a {\it midplane} temperature of 
12,000~K at the shock location (as expected for the ``hot'' spot), 
we obtain $H/r=c_s/V_K=0.02$, and $H_0/H= 4$.  
For comparison,  
\citet{lub89} chose $H \sim 0.01a$, where $a$ is the binary separation,  
which for a disk of diameter $R_{\rm disk}=0.3a$, gives 
$H/r \approx 0.03$ at the edge disk; in that case $H_0/H=2.67$. 
The disk thickness $H$ does not enter the simulations anywhere, except
for the constrain $H_0 > H$. 
The particles are deflected there from 
an initial height $z_i = (1- i/90) \times H_0$. Hence, the thickness $H_0$ 
only gives an estimate of the initial height
from where the $L1$ stream is deflected. 
It is also possible that the 
impact of the stream on the rim of the disk causes the disk to thicken
there, due, for example, to shock heating near the midplane of the disk.
In that case our isothermal assumption might still hold away from the
mid-disk plane, which is from where the particles are launched.

\paragraph{Outer Disk Radius.} 
From our modeling, the actual location of the outer disk radius is 
within about $H_0$ from $R_{\rm disk}$, since this defines
the region  from where the stream material is launched.
We therefore choose $H_0/2$ as the error in the radius of the disk. 
Since $H_0/r  =0.08$, we take $0.04 \times 0.29a \approx 0.012a$ as a first
error.
An additional error of $0.05 R_L = 0.021a$ is introduced as the step
by which the outer disk radius was increased from simulation to simulation.  
As a consequence, 
the outer disk radius we obtained is $0.29a \pm 0.023a$, which  
is smaller than the tidal cutoff radius ($0.33a$; e.g.  \citet{lin09}). 
For a larger disk outer radius, the stream has less velocity,  
it veils the inner disk and impacts the disk at a later phase, and
vice versa \citep[see also][]{kun01}. The outer radius was obtained by fitting 
the phase of maximum velocity offset, and it agrees with the 
theoretical expectation. 

\paragraph{ Inclination.} 
We obtained an inclination of $66^{\circ} \pm 1^{\circ}$, which is 
consistent with the value derived by \citet{lon99,lon06} ($67^{\circ}
\pm 3^{\circ}$)
but noticeably lower than that in \citet{und06} ($72^{\circ}$). 
The (relatively) lower inclination we found is needed to decrease the velocity
maximum and its phase. The simulations were run in incremental steps
of $1^{\circ}$ in $i$ and we therefore take the error to be $1^{\circ}$.  

\paragraph{Launching Angle $\theta$ \& Further Hydrodynamical Considerations.} 
Each simulation includes 91 particles, with launching angle 
$\theta = 0^{\circ}, 1^{\circ}, 2^{\circ}, .., 89^{\circ}, 90^{\circ}$.  
And for each simulation we selected the particles agreeing with the
data. The model that agreed best with the U Gem data included
particles with $\theta < 44^{\circ}$, i.e. $43^{\circ}$ was the maximum
launching angle. 
While it may seem that adjusting $\theta$
in the model is an empirical fit, deflection of hypersonic flow past 
a blunt object is actually limited to a maximum angle $\theta_{\rm max}$
(see subsection 4.1). The fact that we obtained a 
maximum deflection angle is therefore
consistent with the basic hydrodynamical assumption.   
Including particles with a deflection angle $\theta > 43^{\circ}$ causes
the model to depart from the data. 
Furthermore, 
particles that have larger deflection angle 
(say, as $\theta \rightarrow  90^{\circ}$) 
represent the stream material hitting the disk near its denser midplane 
region, where the material strong interaction with the disk prevents it 
from being launched (i.e. the 
deflection/launching assumption is not valid anymore, as pressure effects
due to heating may take place, the vertical velocity decreases to zero, 
and the material is being dragged into a Keplerian orbit with the
disk material). 

\paragraph{ Veiled Inner Region.} 
By fitting the data,
we obtained a veiling region with a radius $R_v=0.025a$, which translates into 
$\sim 1/12$ of the outer radius of the disk or 8 times the WD radius ($8R_{\rm wd}$). 
For a $\sim 1 M_{\odot}$ WD accreting at a mass accretion of near 
$10^{-8}M_{\odot}$/yr \citep{fro01}, 
the veiled inner disk, with $R < 8 R_{\rm wd}$, has a temperature 
of $T \sim 30,000$~K and larger (the standard disk model, e.g.
\citet{pri81}). This is consistent with that region being the
main FUV component in the {\it FUSE} spectral range.  
The outer disk with $T< 20,000$~K ($r> 16R_{\rm wd}$) does not contribute
much flux in the {\it FUSE} range.  
And the disk region with $T \approx 25,000 \pm 5,000$~K (contained 
within $16> r/R_{\rm wd}> 8$) 
contributes less than 1/5 of the hotter inner region ($r/R_{\rm wd} < 8$).  
The size of the veiled region  we obtained is therefore consistent 
with being the main disk region contributing to the FUSE spectrum.  

\paragraph{ The $c_s^1 \beta$ Term.}
In section 4.2, we set the component $c_s^1 \beta$
of the downstream velocity perpendicular
to the shock surface to zero , by assuming that in the isothermal
case it is negligible (equation 10).  
We ran simulations taking that term into consideration
(equation 9), and obtained a difference of
$1^{\circ}$ in the launching angle $\theta$, 
which produces the same results within the error bars.

\paragraph{The Choice of the Function $\alpha (\theta)$.} 
We also checked different functions $\alpha (\theta)$ (equations 14 \& 15)
and found that the exact form of the function had no noticeable
effect on the results. 
 
\paragraph{ Isothermal Assumption.}
The assumption of an isothermal flow is explicitly made in 
equation 10 (i.e. assuming $c_s^1 = c_s^0$), but it is also implicitly  
made in our most basis assumption that the stream-disk interaction
can be represented as a hypersonic flow past a blunt body.   
In the flow past a blunt body we assumed that the stream is deflected
due to the presence of the shock (or `solid' disk surface), rather than due 
to the adiabatic expansion of the heated gas. We expect this assumption
to break down near the disk midplane where the shock is the strongest and
heating is likely to occur. However, that portion of the flow does not
contribute to the computational data since the launching angle occurs
for the upper portion of our simulated bow shock with $\theta < 44^{\circ}$. 
This validates the isothermal assumption as self-consistent.   

\paragraph{Stream Density and Depth of the Absorption Lines.} 
The ballistic trajectories (particle paths) define a strip 
(i.e. a 2D surface like a curtain) that masks the inner disk, 
and the (projected) density of the trajectory ``lines'' is a likely 
indicator of the relative projected thickness of the stream. 
However, as we do not take the pressure effects into account, 
the vertical and lateral expansion of the stream, that is  
likely decreasing its density and turning the strip 
into a three-dimensional conduit, is ignored. Our simulations
also do not include any assumption about the actual mass accretion rate.   
In other words, we did not make any theoretical attempt to derive the stream 
density, and, as a consequence, we cannot predict whether the stream 
has the adequate density to produce the depth of the observed absorption 
lines.  Instead, we assume {\it a priori} that the absorption lines could be
forming in the stream and we check our assumption {\it a posteriori}  
by fitting the velocity offsets of the absorption 
lines with the projected radial stream velocity.  

\paragraph{Full Phase Dependence Absorption.}  

The data reveal absorption lines increasing in depth 
and the sudden appearance of low-ionization lines at orbital phase
$\sim 0.70 \pm 0.15$ \citep{fro01,lon06,god17}, 
which are the same orbital phases
as the X-ray and EUV light-curve dips seen in U Gem in both outburst
and quiescence \citep{mas88,lon96,szk96}. Our ballistic trajectory 
modeling clearly shows how the
$L1$ stream can veil the inner disk and WD and explains the change
in the absorption lines as a function of the orbital phase,
where the exact {\it veiling phase} depends on the size of the disk. 
The increase in the depth of the absorption lines, however,  
has also been observed around orbital phase 0.2-0.35
\citep{lon06,god17}, and \citet{kun01} had shown, using smoothed
particle hydrodynamics simulations,
that parts of the overflowing stream can bounce off the disk face after
hitting it at orbital phase 0.5, creating an additional absorption region
observed around orbital phase 0.2-0.35 \citep{fro01,lon06,god17}. 
We therefore decided to run the simulations further by allowing the particles
to bounce off the disk face. The {\it reflected} particles 
then continue on a trajectory
reaching a maximum altitude (z) and distance (R) from the WD around phase 
0.20-0.25 (see Fig.13 upper left), and then fall back toward the disk
(Fig.13 upper right). There the particles continue on a trajectory  
that either intersects the disk a second time or moves {\it below} 
to the {\it other} side of the disk. We stopped the simulations
before this happens. We then draw, in Fig.13 lower panel, 
the projected radial velocity of the
particles veiling the inner disk as a function of the phase 
for comparison with the observed velocity offsets of the metal
absorption lines. The reflected ballistic trajectory model is able to reproduce
the phase (0.2-0.35) at which    the secondary absorption lines are observed. 
In addition, the particle radial velocity agrees with the velocity 
offsets where there are actual data. In the regions where the stream
only partially veils the inner disk (in blue in Fig.13, lower panel),
we do not expect the absorption-line velocity to be significantly affected 
by the stream. This is especially the case around phase 0.6.
The reflected stream spreads more than the incident stream and one might
therefore expect that the veiling curtain is thinner around phase 0.2-0.35
than around phase 0.7-0.8.  

\clearpage 

\begin{figure}[b!]
\vspace{-4.cm} 
\gridline{ 
           \fig{f13a.eps}{0.5\textwidth}{}
           \fig{f13b.eps}{0.5\textwidth}{}
          }
\vspace{-3.0cm} 
\gridline{ 
           \fig{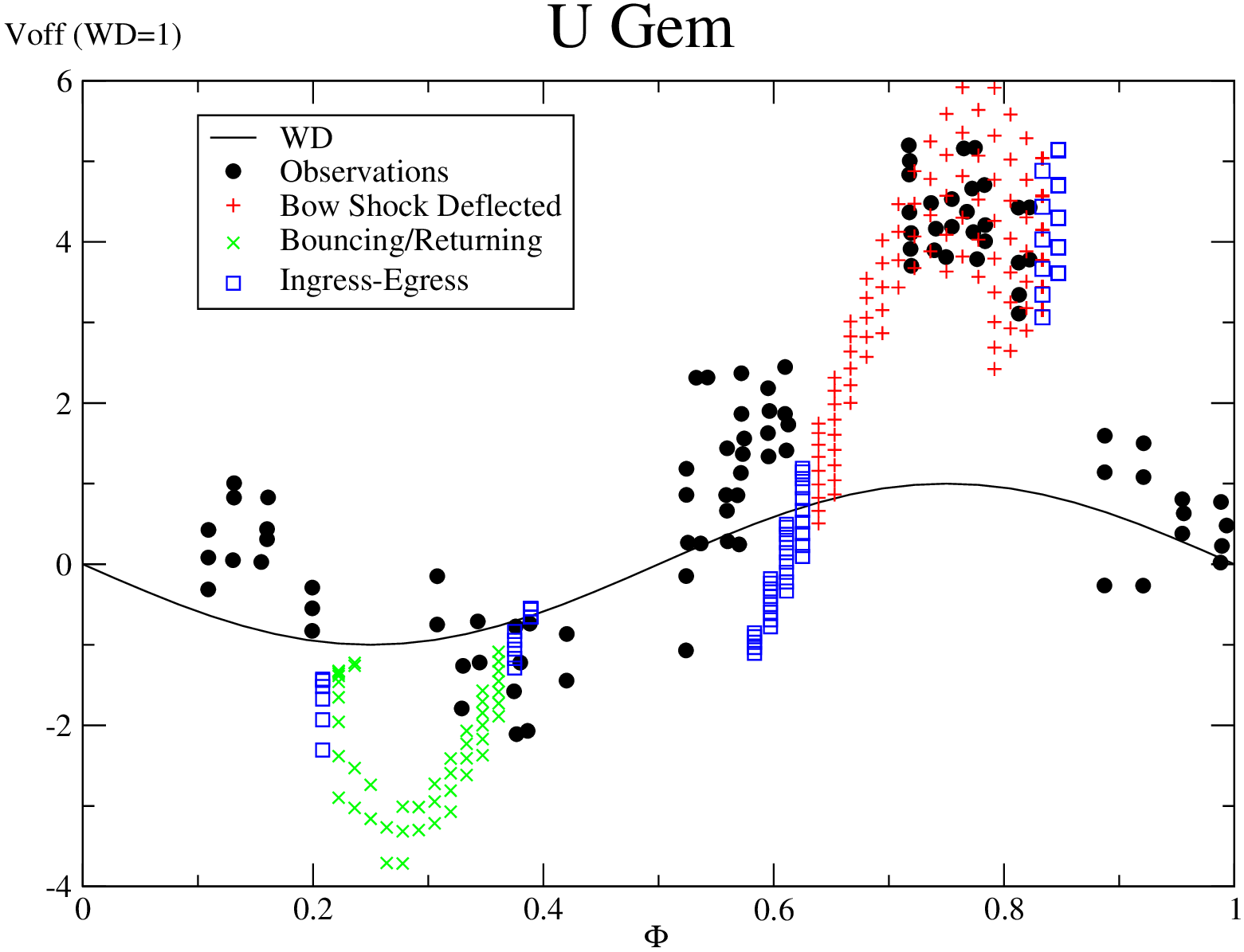}{0.7\textwidth}{}
          }
\vspace{-2.cm} 
\caption{
The deflected incident $L1$ stream is allowed to bounce back off the
disk face (near phase 0.5). The simulation is stopped as the stream
continues toward phase 0.1, before it either bounces back off the disk 
a second time or reaches the other face of the disk.
The top two panels show the ballistic trajectories in the physical
space: face-on (left) and in perspective at phase 0.0 (right).
The lower panel show the resulting projected
(toward the observer) radial velocity as a function of the phase
(as in Figs.11 \& 12). We have now added the theoretical data points
for the bouncing stream (green crosses). The stream covering less
than $\sim 50$\% of the inner disk (at ingress and egress) has been
marked with blue squares.   
}
\end{figure} 

\clearpage

\section{{\bf Summary \& Conclusion}} 
 
Within the context of the Roche model in the restricted 
three-body problem, we model the interaction of the $L1$ stream 
with the accretion disk using ballistic trajectories,  
assuming that the stream-disk interaction can be represented 
as a hypersonic flow past a blunt object. 
Assuming that pressure effects are negligible and that the flow
is isothermal, the bow shock 
deflects the stream material and launches it into ballistic trajectories.  
The $L1$ stream material veils the inner disk and is suspected to 
be responsible for affecting absorption lines in spectra
of high inclination interacting binaries around orbital phase $\sim 0.75$. 
Using this formalism, we model the    
velocity offsets of metal absorption lines of U Gem in outburst 
at orbital phase $\sim 0.70-0.85$.  
The simulations match the velocity and phase of the lines  
for the following system parameters:
a mass ratio $q=m_2/m_1 =0.35$, an inclination $i=66\pm 1^{\circ}$,
an outer disk radius $R_{\rm disk}=0.29a \pm 0.023a$, 
a measure of the bow shock vertical thickness $H_0/r=0.08 \pm 0.02$,   
and the portion of the inner disk that 
is veiled by the stream is confined to $r \le R_v = 0.025a \pm 0.005a$,
where $R_v$ is the radius of the inner disk that is veiled by the stream
and $a$ is the binary separation. 

In our ballistic streamline approximation we neglected pressure
effects, and as a consequence, we did not address the width
and thickness of the stream, nor its vertical density structure
and scale height. Therefore, our model cannot predict whether the stream 
has the adequate density to produce the depth of the observed absorption 
lines.

Because of all the simplifications and assumptions in 
our model, it is likely that the actual values of the outer disk radius, 
the thickness of the bow shock, and the size of the veiled inner disk 
are in fact different than derived here. 
It is also very likely that, during U Gem outburst, $\dot{M}$, $R_{\rm disk}$,
$H_0$ and $R_v$, all vary with time.    
In spite of these, our results provide increased evidence for the
stream overflow model to explain the velocity offsets of the absorption
lines. 

Stream-disk overflow is possibly a prominent and important feature in 
most systems with Roche lobe overflow with a disk, as 
a substantial part of the accretion stream can overflow the disk in
many different systems, in some cases with matter settling around the 
circularization radius \citep{kun01}. 
Observations of semi-detached interacting binaries indeed 
reveal orbital modulation 
in the optical, ultraviolet, and X-ray bands consistent with 
the presence of $L1$ stream material overflowing the disk edge.   
 
We suggest ballistic trajectories as a computational tool  
for the analysis of the absorption-line orbital variability  
in semi-detached binaries. This method could also be used 
with the limitations mentioned above, 
to assess or confirm the 
system parameters such as the mass ratio, inclination, and disk outer radius.  
This formalism can be applied to systems under certain restrictions,
namely, when the launching of the material is occurring in a shocked region
of the stream-disk interaction that is isothermal and where pressure 
and adiabatic effects can be neglected.  

\acknowledgments

It is a pleasure to thank Stephen (Steve) Lubow for a recent discussion
and the interest he shows in our research, 
and William (Bill) P. Blair for his 
kind hospitality at the Henry A. Rowland Department of Physics and 
Astronomy at the Johns Hopkins University, Baltimore, Maryland, USA. 
Cynthia Froning kindly agreed that we use her results digitally extracted
from Fig.7 in \citet{fro01}. 
Support for this work was provided by the National
Aeronautics and Space Administration (NASA) under Grant number 
NNX17AF36G issued through the Office of Astrophysics Data Analysis
Program (ADAP) to Villanova University.

\facilities{{\it IUE}, {\it FUSE}, {\it HST}(COS), {\it HST}(STIS)} 

\software{IRAF \citep{tod86,tod93}, PGPLOT (Tim Pearson, 1995), 
Cygwin-X 
({\tt \url{https://x.cygwin.com/}}), XV (John Bradley, 1998).}   

\begin{center}
\bf{ORCID iDs} 
\end{center} 
Patrick Godon {\tt \url{https://orcid.org/0000-0002-4806-5319}}


\begin{thebibliography}{}

\bibitem[Agafonov et al.(2009)Agafonov, Sharova, \&  Richards]{aga09}
Agafonov, M., Sharova, O., \&  Richards, M., 2009, \apj, 690, 1730 

\bibitem[Armitage \&  Livio(1996)]{arm96}
Armitage, P.J., \&  Livio, M. 1996, \apj, 470, 1024 

\bibitem[Armitage \&  Livio(1998)]{arm98}
Armitage, P.J., \&  Livio, M. 1998, \apj, 493, 898  

\bibitem[Beuermann \&  Thomas(1990)]{beu90}
Beuermann, K., \&  Thomas, H.-C. 1990, \aap, 230, 326 

\bibitem[Billington et al.(1996)]{bil96}
Billington, I., Marsh, R.T., Horne, K., Cheng, F.H., 
Thomas, G., Bruch, A., O'Donoghue, D., \&  Eracleous, M. 1996, 
\mnras, 279, 1274  

\bibitem[Bisikalo et al.(2000)]{bis00}
Bisikalo, D.V., Boyarchuk, A.A., Kuznetsov, O.A., 
\& Chechetkin, V.M. 2000, Astron.Report, 44, 26 
(from Russian: Astro.Zhurnal 77, 2000, 31) 

\bibitem[Bisikalo et al.(2003)]{bis03}
Bisikalo, D.V., Boyarchuk, A.A., Kaygorodov, P.V., \&  Kuznetsov, O.A.,
2003, Astron.Rep., 47, 809 (from Russian: Astro.Zhurnal 80, 2003, 879) 

\bibitem[Bisikalo et al.(2005)]{bis05}
Bisikalo, D.V., Kaygorodov, P.V., Boyarchuk, A.A., 
\&  Kuznetsov, O.A. 2005, Astron.Rep.49, 701 (originally
in Russian: Astron.Zhurnal vol. 82, N.9, 2005, pp. 788-796)  

\bibitem[Blondin(1998)]{blo98}
Blondin, J.M. 1998, API Conf.Proc. 431, 309 

\bibitem[Csizmadia et al.(2008)]{csi08} 
Csizmadia, Sz., Nagy, Zs., Borkovits, T., Heged\"us, T., Biro, I.B.,
\&  Kiss, Z.T. 2008, AN 329, 39 

\bibitem[Dai \& Qian(2009)]{dai09}
Dai, Z., \&  Qian, S. 2009, ApSS, 321, 91 


\bibitem[Dhillon et al.(1997)Dhillon, Marsh, \&  Jones]{dhi97}
Dhillon, V.S., Marsh, T.R., \&  Jones, D.P.H. 1997, \mnras, 291, 694 

\bibitem[Echevarr\'ia et al.(2007)Echevarr\'ia, de la Fuente, \&  Costero]{ech07}
Ecchevarr\'ia, J., de la Fuente, E., \&  Costero, R. 2007, \apj, 134, 262 

\bibitem[Frank et al.(1987)Frank, King, \&  Lasota]{fra87}
Frank, J., King, A.R., \&  Lasota, J.P. 1987, \aap, 178, 137 

\bibitem[Froning et al.(2001)]{fro01}
Froning, C.S., Long, K.S., Drew, J.E., Knigge, C., 
\&  Progra, D. 2001, \apj, 562, 963 

\bibitem[Godon et al.(2017)]{god17} 
Godon, P., Shara, M.M., Sion, E.M., \&  Zurek, D. 2017, \apj, 850, 146 

\bibitem[Godon et al.(2009)]{god09}
Godon, P., Sion, E.M., Barrett, P.E., \&  Linnell, A.P. 2009, \apj, 699, 1229  

\bibitem[Godon et al.(2006)]{god06}
Godon, P., Sion, E.M., Cheng, F., Long, K.S., G\"ansicke, B.T., et al. 2006, \apj, 642, 1018  

\bibitem[Hack \&  La Dous(1993)]{hac93} 
Hack, M., \&  La Dous, C. 1993, {\it Cataclysmic Variables \& 
Related Objects}, (NASA SP-507)/US Gov.Printing Office 

\bibitem[Harrison et al.(2004)]{har04} 
Harrison, T.E., Johnson, J.J., MaArthur, B.E., Benedict, G.F., 
Szkody, P., Howell, S.B., \&  Gelino, D.M. 2004, \aj, 127, 460 

\bibitem[Hartley et al.(2002)]{har02}
Hartley, L.E., Drew, J.E., Long, K.S., Knigge, C., \& Proga, D. 2002,
\mnras, 332, 127 

\bibitem[Hellier(1996)]{hel96}
Hellier, C. 1996, \apj, 471, 949 

\bibitem[Hellier \&  Robbinson(1994)]{hel94} 
Hellier, C., \&  Robbinson, E.L. 1994, \apj, 431 L107 

\bibitem[Kopal (1959)]{kop59}
Kopal, Z. 1959, {\it Close Binary Systems}, New York: Wiley 

\bibitem[Kruszewski(1966)]{kru66}
Kruszewski, A. 1966, {\it Adv.Astron.\& Astrop.}, 4, 233  

\bibitem[Krzeminski(1965)]{krz65}
Krzeminski, W. 1965, \apj, 142, 1051 

\bibitem[Kuiper(1941)]{kui41}
Kuiper, B.P. 1941, \apj, 93, 133  

\bibitem[Kunze et al.(2001)Kunze, Speith, \&  Hessman]{kun01} 
Kunze, S., Speith, R., \&  Hessman, F.V. 2001, \mnras, 322, 499 

\bibitem[Linnell et al.(2007)]{lin07}
Linnell, A.P., Godon, P., Hubeny, I., Sion, E.M., \&  Szkody, P., 
2007, \apj, 662, 1204 

\bibitem[Linnell et al.(2009)]{lin09}
Linnell, A.P., Godon, P., Hubeny, I., Sion, E.M., Szkody, P., \& 
Barrett, P.E.  2009, \apj, 703, 1839 
     

\bibitem[Long \& Gilliland(1999)]{lon99} 
Long, K.S., \& Gilliland, R.L. 1999, \apj, 511, 916 

\bibitem[Long et al.(2006)Long, Brammer, \& Froning]{lon06}
Long, K.S., Brammer, G., \& Froning, C.S. 2006, \apj, 648, 558 

\bibitem[Long et al.(1993)]{lon93}
Long, K.S., Blair, W.P., Bowers, C.W., Davidsen, A.F., Kirss, G.A.,
Sion, E.M., \& Hubeny, I. 1993, \apj, 405, 327 

\bibitem[Long et al.(1996)]{lon96} 
Long, K.S., Mauche, C.W., Raymond, J.C., Szkody, P., \& Mattei, J.A. 
1996, \apj, 469, 841 

\bibitem[Lubow(1989)]{lub89}
Lubow, S.H. 1989, \apj, 340, 1064 

\bibitem[Lubow \& Shu(1975)]{lub75}
Lubow, S.H., \& Shu, F.H. 1975, \apj, 198, 383 

\bibitem[Lubow \& Shu(1976)]{lub76}
Lubow, S.H., \& Shu, F.H. 1976, \apjl, 207, L53 

\bibitem[Marsh(1985)]{mar85}
Marsh, T.R. 1985, Ph.D. thesis, Cambridge University 

\bibitem[Marsh \& Horne(1988)]{mar88}
Marsh, T.R., \& Horne, K. 1988, \mnras, 235, 269  

\bibitem[Marsh et al.(1990)]{mar90}
Marsh, T.R., Horne, K., Schlegel, E.M., Honeycutte, R.K., 
\& Kaitchuck, R.H. 1990, \apj, 364, 637 

\bibitem[Mason et al.(1988)]{mas88}
Mason, K.O., C\'ordova, F.A., Watson, M.G., \& King, A.R. 1988, \mnras, 
232, 779 

\bibitem[Moulton(1914)]{mou14}
Moulton, F.R. 1914, {\it An Introduction to Celestial Mechanics} 
(New York: Macmillan), 2nd revised edition, 
reprinted by Dover Publications, New York 1984 

\bibitem[Murray(1991)]{mur96}
Murray, J.R. 1996, \mnras, 279, 402  

\bibitem[Naylor et al.(2005)]{nay05}
Naylor, T., Allan, A., Long, K.T. 2005, \mnras, 361, 1091 

\bibitem[Nussbaumer \& Orr(1994)]{nus94}
Nussbaumer, H., \& Orr, A., (eds) 1994, {\it Interacting Binaries},
(Berlin: Springer-Verlag)

\bibitem[Paczy\'nski(1971)]{pac71}
Paczy\'nski, B. 1971, \araa, 9, 183  

\bibitem[Peris et al.(2015)]{per15}
Peris, C.S., Vrtilek, S.D., Steiner, J.F., Vrtilek, J.M., Wu, J.,
et al. 2015, \mnras, 449, 1584 

\bibitem[Plummer(1918)]{plu18}
Plummer, H.C. 1918, {\it An Introductory Treatise on 
Dynamical Astronomy} Cambridge University Press,
reprinted by Dover Publications, New York 1960 

\bibitem[Prendergast \& Taam(1974)]{pre74}
Prendergast, K.H., \& Taam, R.E. 1974, \apj, 189, 125  

\bibitem[Press et al.(1992)]{numrec}
Press, W.H., Teukolsky, S.A., Vetterling, W.T., \& Flannery, B.P.,
Numerical Recipes in Fortran 77, The Art of Scientific Computing,
Second Edition, 1992, Cambridge University Press.

\bibitem[Pringle(1981)]{pri81}
Pringle, J.E. 1981, \araa, 19, 137  

\bibitem[Rathakrishnan(2010)]{rat10}
Rathakrishnan, E. 2010, {\it Applied Gas Dynamics}, 
1st Edition, John Wiley \& Sons (Asia), Singapore

\bibitem[Ritter \& Kolb(2003)]{rit03}
Ritter, H., \& Kolb, U. 2003, \aap, 404, 301 
(update RKcat7.24, 2016) 
$<$http://physics.open.ac.uk/RKcat/RKcat\_AA.ps$>$

\bibitem[Schmidtobreick et al.(2003)Schmidtobreick, Tappert, \& Saviane]{sch03} 
Schmidtobreick, L., Tappert, C., \& Saviane, I. 2003, \mnras, 342, 145  

\bibitem[Sion et al.(1998)]{sio98}
Sion, E.M., Cheng, F.H., Szkody, P., Sparks, W., G\"ansicke, B.T., 
Huang, M., \& Mattei, J. 1998, \apj, 496, 449 

\bibitem[Sion et al.(2017)Sion, Godon, \& Jones]{sio17}
Sion, E.M., Godon, P., \& Jones, L. 2017, \aj, 153, 109 

\bibitem[Smak(1971)]{sma71}
Smak, J.I. 1971, Acta Astronomica, 21, 15  

\bibitem[Smak(2001)]{sma01}
Smak, J.I. 2001, Acta Astronomica, 51, 279 

\bibitem[Spruit \& Rutten (1998)]{spr98}
Spruit, H.C., \& Rutten, R.G. 1998, \mnras, 299, 768 

\bibitem[Szkody et al.(1996)]{szk96} 
Szkody, P., Long, K.S., Sion, E.M., \& Raymond, J.C. 1996, \apj, 469, 834 

\bibitem[Tody(1986)]{tod86}
Tody, D., 1986, in Crawford D.L., ed., Society of Photo-Optical Instrumentation
Engineers (SPIE) Conference Series Vol.627, Instrumentation in Astronomy VI.
p. 733 

\bibitem[Tody(1993)]{tod93}
in Hanisch R.J., Brissenden, R.J.V., Barnes, J., eds, Astronomical Society of
the Pacific Conference Series Vol.52, Astronomical Data Analysis Software and
Systems II. p. 173  

\bibitem[Unda-Sanzana et al.(2006)Unda-Sanzana, Marsh, \& Morales-Rueda]{und06}
Unda-Sanzana, E., Marsh, T.R., \&  Morales-Rueda, L. 2006, \mnras, 369, 805 

\bibitem[Vogt et al.(2017)]{vog17}
Vogt, N., Schreiber, M.R., Hambsch, F.-J., Retamales, G., Tappert, C. et al.
2017, \pasp, 129, 4201 

\bibitem[Warner \&  Nather(1971)]{war71} 
Warner, B., \&  Nather, R.E. 1971, \mnras, 152, 219 

\bibitem[Zhang \&  Robinson(1987)]{zha87}
Zhang, E.H., \& Robinson, E.L. 1987, \apj, 321, 813  

\end{thebibliography}
\end{document}